# DNA RENATURATION AT THE WATER-PHENOL INTERFACE

**Running title: Interfacial DNA Renaturation**

**Arach Goldar[§, ‡] * and Jean-Louis Sikorav[§] ***

[§] *Laboratoire de Biophysique de l'ADN, CEA/Saclay, DBJC/SBGM, 91191 Gif-sur-Yvette Cedex, France*

[‡] *Service de Physique des Etats Condensés, CEA/Saclay, DRECAM, 91191 Gif-sur-Yvette Cedex, France*

*Address correspondence to A. Goldar (Telephone: 33 1 69 08 75 98. FAX: 33 1 69 08 43 89. E-mail: agoldar@cea.fr) or J.-L. Sikorav (Telephone: 33 1 69 08 66 43. FAX: 33 1 69 08 43 89. E-mail: jlsikorav@cea.fr)

**Abbreviations**: PERT, phenol emulsion reassociation technique; EDTA, ethylenediaminetetraacetic acid; Tris, tris(hydroxymethl)aminomethane; ssDNA, single-stranded DNA; dsDNA, double-stranded DNA; SSBP, single-stranded DNA binding protein; T4 gp32, gene 32 protein of bacteriophage T4; *Eco* SSB, *Escherichia coli* single-stranded binding protein

**Abstract**

We study DNA adsorption and renaturation in a water-phenol two-phase system, with or without shaking. In very dilute solutions, single-stranded DNA is adsorbed at the interface in a salt-dependent manner. At high salt concentrations the adsorption is irreversible. The adsorption of the single-stranded DNA is specific to phenol and relies on stacking and hydrogen bonding. We establish the interfacial nature of a DNA renaturation at a high salt concentration. In the absence of shaking, this reaction involves an efficient surface diffusion of the single-stranded DNA chains. In the presence of a vigorous shaking, the bimolecular rate of the reaction exceeds the Smoluchowski limit for a three-dimensional diffusion-controlled reaction. DNA renaturation in these conditions is known as the Phenol Emulsion Reassociation Technique or PERT. Our results establish the interfacial nature of PERT. A comparison of this interfacial reaction with other approaches shows that PERT is the most efficient technique and reveals similarities between PERT and the renaturation performed by single-stranded nucleic acid binding proteins. Our results lead to a better understanding of the partitioning of nucleic acids in two-phase systems, and should help design improved extraction procedures for damaged nucleic acids. We present arguments in favor of a role of phenol and water-phenol interface in prebiotic chemistry.

The most efficient renaturation reactions (in the presence of condensing agents or with PERT) occur in heterogeneous systems. This reveals the limitations of homogeneous approaches to the biochemistry of nucleic acids. We propose a heterogeneous approach to overcome the limitations of the homogeneous viewpoint.

# Introduction

In 1977 Kohne, Levison and Byers (1) reported a striking experiment allowing a fast renaturation of complementary nucleic acids. They put denatured nucleic acids in an aqueous solution containing monovalent salts. Enough phenol was then added to obtain a second liquid phase. The constant and vigorous shaking of this biphasic system at room temperature created an instable emulsion in which the fast renaturation of the nucleic acids took place. They called this method the phenol emulsion reassociation technique or PERT. PERT accelerates RNA-RNA as well as RNA-DNA reactions, but is most efficient with DNA-DNA reactions at very low concentrations (1, 2). Under these conditions, the rate of DNA renaturation is independent of the length of the complementary strands, and is "many thousand times faster than under the standard conditions" (1). The rate shows a weak temperature dependence: the rate of renaturation is only 3-4 times faster at 56° C than at room temperature (1). In addition the rate depends on a critical manner on the salt concentration (3). PERT has been used with profit in many laboratories; it has allowed for instance Kunkel and coworkers (4) to isolate the gene responsible for the muscular dystrophy of Duchenne de Boulogne. The mechanism of PERT is still poorly understood. Kohne et al. (1) realized the possible physiological interest of such a system as well as its plausible prebiotic significance, but the lack of understanding of its basic mechanism seems to have precluded further investigations along these lines.

We have undertaken an experimental study of the behavior of nucleic acids in water-phenol two-phase systems, with the initial goal of clarifying the mechanism of PERT. We have hypothesized that PERT involves the coupling between two processes, single-stranded DNA (ssDNA) adsorption and annealing initiated at the water-phenol interface. We have therefore decided to study this reaction by using a decoupling scheme: instead of examining solely the behavior of the two complementary ssDNA chains obtained through the denaturation of a native double-stranded DNA (dsDNA) (1-3), we have chosen to study 1) the behavior of each ssDNA molecule in the absence of their complementary strand (with or without shaking) and 2) the renaturation of the two complementary strands with various stoichiometries (again with or without shaking). Similar decoupling approaches have already been employed for other systems: DNA renaturation coupled with DNA aggregation in the presence of RecA protein (5); DNA renaturation coupled with ssDNA folding (6, 7) and DNA renaturation coupled with DNA aggregation in the presence of spermine (8).

In this work we use this decoupling scheme to investigate the partitioning and renaturation of very dilute ssDNA solutions in a water-phenol two-phase system. We first demonstrate that there is a salt-dependent adsorption of a single-stranded DNA molecule in the absence of its complementary strand at the water-phenol interface. The adsorption is irreversible at high monovalent salt concentrations. In addition to phenol, we test thirteen other organic solvents for their ability to confine the single-stranded DNA at the water-solvent interface. Phenol turns out to be the most efficient solvent. We study the consequences of the gradual addition of an ssDNA in a high salt concentration water-phenol two-phase system where the complementary strand has been previously irreversibly adsorbed. We observe the progressive release of the adsorbed complementary strand in the aqueous phase, and show that this strand has renatured with the added ssDNA. This establishes the interfacial nature of the renaturation reaction in these conditions. We further investigate the effect of shaking on the rate of adsorption and renaturation. In the absence of shaking, adsorption is a slow, monomolecular diffusion-controlled process. The adsorption is completed within a few seconds if the solution is vigorously shaken. The study of DNA renaturation in the absence of shaking shows the existence of a regime where the rate of the reaction depends on the two-dimensional diffusion of the two complementary single-strands. We finally establish the interfacial nature of PERT, and determine the bimolecular rate of renaturation obtained with this technique. The measured value ($4.2 \times 10^{10}$ $M^{-1}$ $s^{-1}$) exceeds the rates obtained using either condensing agents or single-stranded DNA binding proteins (SSBPs).

These experimental studies have progressively led us to ponder on general questions dealing with fundamental and applied biochemistry of nucleic acids. We will list briefly these questions, and examine them in the discussion section:

1) Adsorption and renaturation of DNA in the water-phenol two-phase system. What is the origin of the efficiency of phenol as a single-stranded nucleic acid binding and confining compound? What are the respective roles of hydrophobicity and aromaticity in the interaction of phenol and other simple organic compounds with nucleic acids (9-19)? What is the mechanism of the interfacial renaturation, with or without shaking?

2) Comparison with other renaturing approaches. How does the mechanism of this interfacial renaturation (with or without shaking) compare with the standard thermal renaturation approach (20-24), and with approaches involving condensing agents (3, 8, 25) or single-stranded binding proteins (SSBPs, often referred to as nucleic acid chaperones) (26-34)? What are the consequences of the coupling of the adsorption process with the coil-helix transition for the mechanism of this interfacial renaturation (35, 36)? How does the interaction of phenol compare with the interaction of tyrosine and other aromatic amino acids of SSBPs with single-strand nucleic acids? In the absence of shaking, our results show the existence an efficient two-dimensional diffusion process of the irreversibly adsorbed ssDNA chains. In a general manner, how can one obtain an efficient surface diffusion for nucleic acids?

3) Implications for the partitioning of nucleic acids in water-phenol and other two-phase systems. The purification of nucleic acids in two-phase systems (water-chloroform, water-phenol or water two-phase systems) is a common technique of molecular biology (37-45). The partitioning of nucleic acids in water-phenol systems depends on numerous parameters: nucleic acid structure and conformation (double-stranded, single-stranded, denatured, supercoiled), nucleic acid concentration and sequence, pH, temperature, and salt concentration (40, 46-51). Losses of nucleic acids due to an "aggregation" at the water-chloroform or water-phenol interface or a transfer to the organic phase have been reported (51-55): what is the role of surface phenomena in the partitioning of single and double-stranded nucleic acids?

4) Implications for the evolution of biological chemistry. Phenol is known to be a plausible prebiotic compound (see for instance (56) and further references therein). Furthermore, phenol is a very efficient nucleic acid binding and confining compound. This raises the possibility that phenol was an active compound in an early nucleic acid world (57-59). What could have been the role of phenol and phenolic compounds in prebiotic chemistry and in the early stages of life? Surface phenomena, in particular adsorption on solid surfaces such as clays or other mineral surfaces are thought to have had an important role in prebiotic chemistry (60-64). Can mechanisms involving an adsorption at a liquid-liquid interface have also contributed to prebiotic chemistry?

# Materials and methods

## Materials

### *DNA*

Two synthetic complementary 118 base-long single-stranded DNAs denoted 118+ (MW = 36318.3 Da) and 118- (MW = 36477.4 Da) were obtained from Eurogentec. The sequence of 118+ corresponds to the sequence of the nucleotides 4760-4877 of the plus strand of $\phi$X 174 (65) (positions 4758 and 4876 correspond to restriction sites for the enzyme Hae III).

### *Phenol and other organic solvents*

Phenol (Molecular Biology grade) was obtained from Sigma and stored at –20 °C. We used a modification of the protocol described in (43) to prepare a water-phenol solution with a phenolic pH greater than 7.8. Phenol was melted at 68 °C, and an equal volume of 0.5 M Tris-HCl (pH = 8) at room temperature was added to it. The mixture was shaken at room temperature for 15 minutes and left at rest to separate the two phases. The organic phase was then collected using a separating funnel. This process was repeated with an equal volume of 0.1 M Tris-HCl (pH = 8), until the pH of the phenolic phase (measured using a pH meter calibrated for aqueous solutions) exceeds 7.8 (this required at least two more extractions). When the proper pH was obtained (usually 7.85), the aqueous phase was removed and replaced by one volume of 0.1 M Tris-HCl (pH = 8). This solution was kept at 4 °C and used within a month. We did not add 8-hydroxyquinoline (39), because the presence of this compound modifies the partitioning of ssDNA in the water-phenol two-phase system (data not shown).

Dodecane, carbon tetrachloride, chloroform and benzylalcohol were obtained from Merck; Chlorobenzene, fluorobenzene, iodobenzene and guaiacol (2-methoxyphenol) from Aldrich;

Aniline and bromobenzene from Fluka; 1-Butanol from Prolabo; dichloromethane from SdS, and toluene from Sigma.

*Plasticware*

The experiments described in this work were performed at extremely low ssDNA concentrations (typically with a chain concentration in the 1-100 picomolar range). Low DNA concentrations together with high salt concentrations can lead to a significant adsorption on the walls of the tube if no precautions are taken. To prevent this effect, all experiments were done in 1.7 ml low-binding microcentrifuge tubes (Marsh Biomedical Products) coated with a methyl brush as follows: the tubes were filled with a solution of 2% dimethyldichlorosilane in (1, 1, 1)-trichloroethane (BDH). After at least 6 hours of incubation, the silane solution was removed and the tubes were thoroughly rinsed with a TE buffer (10 mM Tris HCl (pH = 7.5), 1 mM EDTA). In the water-phenol system, the adsorption of ssDNA on the walls of the tubes (determined as explained below) is always less than 6 % of the total amount of ssDNA present in the tube at all salt concentrations between 0 and 3 M.

*DNA labeling*

The ssDNA 118+ and 118- were 5'-end labeled with T4 polynucleotide kinase using $\gamma^{32}$P-ATP (> 5000 Ci/mmol, AmershamPharmaciaBiotech) to a specific radioactivity of about $10^8$ cpm/µg. The labeled ssDNAs were separated from unincorporated $\gamma^{32}$P-ATP by gel filtration on a Sephadex-G50 column (Nick Column, Pharmacia) equilibrated in a TE buffer (10 mM Tris HCl (pH = 7.5), 1 mM EDTA). The amount of unincorporated $\gamma^{32}$P-ATP still present in the solution of labeled ssDNA after the gel filtration was determined by an autoradiography of a sample submitted to a separation by gel electrophoresis as described below. The unincorporated $\gamma^{32}$P-ATP accounted for 5 % or less of the total radioactivity present in the solution of labeled ssDNA.

*Partitioning experiments by vortexing*

Unless stated otherwise, all partitioning experiments involving vortexing were performed as follows. The labeled ssDNA (160 µl containing the labeled ssDNA at a chain concentration of 125 pM (about 0.58 ng) in a TE buffer with various concentrations of NaCl) was vortexed for 20 s at 2400 rpm (rounds per minute). The emulsion was then centrifuged at 15000 × g for 1 min. After centrifugation, 100 µl of the aqueous phase and 10 µl of the organic phase were removed and their radioactivity was determined by scintillation counting. Direct Cerenkov counting was generally used for both types of samples (aliquots of aqueous or organic phases). We checked the reliability of these measurements for the organic phases by adding a scintillation cocktail (Pico-Fluor 40, Packard) to the aliquot and counting again. For phenol only, we also determined the amount of radioactivity adsorbed on the tube, by adding 1 ml of a TE solution containing the same NaCl concentration to the tube containing the rest of the two-phase mixture (60 µl of aqueous phase and 30 µl of organic phase): this yields a homogeneous liquid phase. The tube was vortexed for 20 s, the liquid phase removed and the radioactivity still bound to the walls of the tube was determined.

*Renaturation experiments*

The products of the renaturation experiments were separated by gel electrophoresis on a 15 % polyacrylamide gel. The gel was dried and the quantities of single and double-stranded DNA as well as of unincorporated $\gamma^{32}$P-ATP were determined using a Phosphor Imager (Molecular Dynamics).

## Results

*Partitioning of ssDNA in the water-phenol system*

Figure 1 shows the partitioning of the $^{32}P$ labeled ssDNA 118- in the water-phenol two-phase system as function of NaCl concentration. Below 0.3 M NaCl the ssDNA remains in the aqueous phase; it is almost completely removed from this phase at higher salt concentrations. Only a small amount of ssDNA is transferred to the organic phase (typically 10 % or less of the radioactive material) or is found adsorbed on the walls of the tubes (typically 6 % or less of the radioactive material). The amount of radioactivity present in these three phases (aqueous phase, organic phase and surface of the tube) represents only 20% of the total radioactivity above 0.5 M NaCl. Conversely, the difference between the total radioactivity and this amount (represented by full squares in figure 1 accounts for 80% of the total radioactivity. To localize this missing radioactivity we performed a similar partitioning experiment at 0.85 M NaCl in a bigger (15 ml) tube also treated with dimethyldichlorosilane beforehand, and obtained an overnight autoradiography of the tube (an experiment suggested to us by Gilbert Zalcser). Figure 2 shows a picture of the tube, and the autoradiography of a portion of it. There is a peak located at the water-phenol interface (Figure. 2b), which accounts for the missing radioactivity (there is no radioactivity at the air-water interface). We conclude that above 0.5 M NaCl the missing radioactivity is located at the water-phenol interface.

The experiments described below in the water-phenol system have been performed at 0.85 M NaCl. We therefore characterized in greater detail the partitioning of ssDNA at this concentration by repeating this experiment 10 times: 1) 80 ± 8% of the ssDNA is adsorbed the water-phenol interface; 2) 6 ± 1% of the ssDNA is transferred in the phenolic phase; 3) 3-5 % the ssDNA is adsorbed on the walls of the tubes; 4) 3-5% of the total radioactivity is present in the aqueous phase. This radioactivity does not correspond to ssDNA but to $\gamma^{32}$P-ATP (aliquots of the aqueous phase run on a polyacrylamide gel never show the ssDNA band but only the $\gamma^{32}$P-ATP band; data not shown). We also obtained similar results in experiments with the complementary ssDNA 118+ (data not shown).

The reversibility of the adsorption as a function of NaCl concentration was determined in the following manner. We first adsorbed the $^{32}$P labeled 118+ by vortexing and centrifugation as described above in the presence of different NaCl concentration between 0 and 1M NaCl. We measured the amount of radioactivity present in the aqueous phase (figure 3, open circles). We then prepare 10 different tubes where the $^{32}$P labeled 118+ was adsorbed in the presence of 1 M NaCl. We removed 155 µl of the aqueous phase, and replace it by 155 µl of a phenol-saturated TE solution containing different NaCl concentrations. We again vortexed and centrifuged the tubes and determined the amount of radioactivity present in the aqueous phase. Figure 3 (full square) shows the result of this desorption experiment. The adsorbed 118+ is completed desorbed at low salt (0.1 M and below). One observes a hysteresis between 0.2 and 0.6 M NaCl. The differential partitioning of the ions between the two phases could account for this hysteresis**.** The adsorption is irreversible at high salt above 0.6 M. We have checked this irreversibility at 0.85 M by repeating several times the desorption experiment at this salt concentration: we observed no release of the adsorbed DNA.

*Partitioning of ssDNA using other organic solvents*

We performed similar partitioning experiments using 13 other organic solvents that are only partially miscible with water. The results are summarized in Table 1. The solvents were divided into three classes: simple hydrophobic solvents, aromatic solvents and primary alcohols. Five solvents are unable to remove the ssDNA from the aqueous phase, even at 3 M

NaCl (dodecane, toluene, carbon tetrachloride, butanol and aniline). Dichloromethane and chloroform begin to adsorb ssDNA above 2.5 M NaCl only. The other solvents are all planar aromatic compounds. The best solvents are fluorobenzene, guaiacol and phenol. Figure 4 shows that they all start to remove the ssDNA from the aqueous phase at low NaCl concentrations (above 0.3 M NaCl for fluorobenzene and phenol, and 0.5 for guaiacol). However, phenol is the most efficient solvent, allowing a complete adsorption above 0.5 M, a salt concentration where about 80% of the ssDNA is still in the aqueous phase in the presence of fluorobenzene or guaiacol.

*Renaturation and transfer to the aqueous phase of an adsorbed ssDNA in the presence of its complementary strand*

As shown before, at 0.85 M NaCl, 80 % of the ssDNA (118+ or 118-) is irreversibly adsorbed at the water-phenol interface. We performed the following experiment at this salt concentration. We first adsorbed the $^{32}P$ labeled 118- by vortexing and centrifugation as described above, and then removed 155 µl of the aqueous phase. We replaced it by 155 µl of a TE solution containing the same concentrations of phenol and NaCl, and various amounts of the unlabeled complementary strand 118+. We again vortexed and centrifuged the tube, and then determined the amount of radioactivity present in the aqueous phase. Figure 5a shows that the addition of increasing amounts of the complementary strand 118+ leads to the progressive release of the adsorbed 118- in the aqueous phase. Up to 70 ± 10 % of the 118- can be released; this plateau value is obtained when a concentration of 4 ng/ml of 118+ is reached, corresponding to the amount of 118- present in the tube. The release therefore involves the formation of a 1-1 complex. The release is specific to the complementary strand, since it is not observed when similar amounts of the same strand 118- (instead of the complementary strand) are added in the aqueous phase (Figure 3, open triangles). We analyzed the released material by gel electrophoresis: it consists solely of a renatured double-stranded DNA (data not shown). Since the adsorption at the water-phenol interface is irreversible for each of the two complementary ssDNA taken separately in our experimental conditions, the renaturation reaction must have been initiated at the interface. Furthermore, the double-stranded DNA that is the outcome of this reaction is not adsorbed at the interface (it partitions entirely in the aqueous phase in our experimental conditions). Thus, as soon as a double-stranded DNA segment is nucleated, it is expelled from the interfacial region. Figure 5b shows that there is initially a strict linear relation between the amount of added 118+ and the amount of released 118-, which confirms that the release involves a stoichiometric process. The slope of the curve in figure 5b is equal to 0.63 ± 0.01. This value corresponds to the product of the two fractions of adsorbed ssDNA ($0.8 \times 0.8 = 0.64$). This confirms the strict interfacial nature of the reaction: only the adsorbed chains can participate in the reaction. The height of the plateau (70 ± 10 %) corresponds to the complete release of 118- by the renaturation reaction.

*Effects of vortexing on adsorption and renaturation*

*a. Adsorption without vortexing*

Figure 6a shows the percentage of radioactive material removed from the aqueous phase as a function of time in the absence of vortexing. The inset in figure 6a further shows that this percentage increases at first linearly with the square root of time, up to about $1.5 \times 10^4$ s. This dependency is suggestive of an irreversible, diffusion-controlled process (*66-68*). We will model the kinetic of ssDNA adsorption assuming that the adsorption is diffusion limited. At time $t = 0$ the bulk number of ssDNA chains is $N_0$ ($N_0 = C_0 \times V$). The chain concentration $C_0$ is supposed to remain constant close to the air-water interface. This assumption, known as a semi-infinite approximation, will be justified a posteriori. The number $N(t)$ of ssDNA chains adsorbed at time $t$ at the water-phenol interface is obtained by solving a one-dimensional diffusion equation with boundary conditions pertaining to the conical geometry of our system (figure 6b). Under these conditions the rate of adsorption is given by:

$$\left.\frac{dN(t,z)}{dt}\right|_{z=0} = 4\sqrt{\pi}\sin\left(\frac{\pi}{12}\right)\left(ah+\frac{h^2}{3}\right)C_0\sqrt{D_{3d}}\frac{1}{\sqrt{t}} \qquad 1$$

Where $D_{3d}$ is the three-dimensional diffusion coefficient of the ssDNA chain 118-, $h$ is the height of the water column, $a$ is the distance between the water-phenol interface and the apex of the cone, and the interface lies at $z = 0$ (see figure 6b). By integrating both sides of this equation and using the boundary condition $N(t = 0) = 0$, we obtain the following expression for the number of adsorbed ssDNA chains as a function of time.

$$N(t) = 8\sqrt{\pi}\sin\left(\frac{\pi}{12}\right)\left(ah+\frac{h^2}{3}\right)C_0\sqrt{D_{3d}}\sqrt{t} \qquad 2$$

In our experiments, 5% of the total amount of radioactivity corresponds to $\gamma^{32}$P-ATP, which always remains in the aqueous phase. In order to model the experimental results, a constant $A$ should therefore be added to the right hand side of equation 2.

$$\frac{N(t)}{N_0} = \frac{8\sqrt{\pi}\sin\left(\frac{\pi}{12}\right)\left(ah+\frac{h^2}{3}\right)\sqrt{D_{3d}}}{V}\sqrt{t} + A \qquad 3$$

Equation 3 has been used to fit the first regime of adsorption (Figure 4.a). The values of the fitting parameters are:

| Parameters | $D_{3d}$ (cm$^2$.s$^{-1}$) | A |
|---|---|---|
| Values | $2.4 \times 10^{-7} \pm 9.6 \times 10^{-9}$ | $4.6 \times 10^{-2} \pm 8.2 \times 10^{-3}$ |

The value of the constant $A$ is in good agreement with the value of 5% determined above. From the value obtained for $D_{3d}$ we can estimate the characteristic time $t_h$ required to explore the height $h$ of the aqueous phase using the relation $h^2 = 6D_{3d}t_h$. We obtain a time $t_h$ of about $4.4 \times 10^5$ s. The time $t_h$ corresponds to time required to deplete the air-water interface. Since the fitted experimental data have been obtained on a much shorter time scale, the semi-infinite approximation is indeed valid.

*b. Adsorption with vortexing*

The effects of the duration of vortexing and of the vortexing velocity on the kinetics of adsorption of the $^{32}P$ labeled ssDNA 118- are shown in figure 7. At the highest vortexing velocity (2400 rpm), a complete adsorption is achieved immediately (that is within the first five seconds). The rate of adsorption increases with the vortexing velocity for lower velocities. The curves suggest the existence of two regimes. At first there is a roughly linear increase of the amount of adsorbed DNA with time. In the second regime the increase is much reduced; the amount of adsorbed ssDNA at this quasi-plateau increases with vortexing velocity.

*c. Renaturation without vortexing*

Figure 8a shows the percentage of renatured DNA as a function of time obtained in the two-phase system in the absence of vortexing (during the renaturation stage). In this experiment, $^{32}P$ labeled ssDNA 118- (125 pM in the aqueous phase) was first adsorbed at the water-phenol interface by vortexing and centrifugation. We then removed 155 µl of the aqueous phase, and replaced it by 155 µl of a TE solution containing the same concentrations of phenol and NaCl, and the $^{32}P$ labeled complementary strand 118+ (also at 125 pM). The two-phase system was left to react without agitation. The reaction was quenched by the addition of enough TE buffer (600 µl TE buffer) to destroy the two-phase system, and an aliquot was analyzed by gel electrophoresis. The rate renaturation is slow, less than 40% of the DNA having renatured within three hours. A comparison of the rate of this reaction with the rate of adsorption observed without agitation (figure 8b) shows that initially the two processes occur at similar rates; later on renaturation becomes the slower of the two processes. This is fully consistent with the conclusion drawn above that renaturation takes place at the interface: the complementary strand must first be adsorbed before it renature. Later on the percentage of renatured DNA increases linearly with the square root of time (Figure 6.b). We now propose a model to explain this dependency.

The renaturation reaction has two kinetics components (Figure 7): The ssDNA must be adsorbed and then it has to search the interface for its complementary strand to renature. At the beginning of the reaction, adsorption and renaturation occur at similar rates. Renaturation is therefore not rate limiting: the concentration of adsorbed 118- is high enough, and when the 118+ reaches the interface it encounters quickly its complementary and forms a dsDNA which is expelled into the aqueous phase. As the reaction proceeds, the concentration of available adsorbed 118- decreases and therefore the rate of the bimolecular reaction between the two complementary strands slows down. In this second regime the rate of the reaction depends on a combination of the diffusion-controlled rate of adsorption and the interfacial recombination. According to this picture, at still longer times, there is a third regime where the rate of reaction is given solely by the rate of recombination at the interface (we do not reach this regime in the present study). The transition between the first and the second regime can be described using the standard terminology of surface chemistry (*69, 70*). When a reaction between two compounds is catalyzed by a surface, it is possible to distinguish two types of processes. In the first process, the reaction occurs between the two adsorbed compounds, and involves a surface diffusion of at least one of them. This is known as a Langmuir-Hinshelwood mechanism. In the second process, there is no need for a surface diffusion: the reaction occurs between the adsorbed compound and the second compound coming from the volume above the surface. The encounter between the two compounds involves only a 3-dimensional diffusion. This is known as an Eley-Rideal mechanism. In our case, the first regime is compatible with an Eley-Rideal mechanism: where 118+ encounters the adsorbed 118- through a 3-dimensional diffusion (path A in figure 9). A surface diffusion step (path B in Figure 9) may be present, but is too fast to be detected. In the second regime, the rate of renaturation becomes slower than the rate of adsorption. This corresponds to a Langmuir-Hinshelwood mechanism (path B in Figure 7).

We will model the recombination reaction at the interface assuming a diffusion-controlled annealing process. We further assume that the 118- and the 118+ ssDNA have the same shape and two-dimensional diffusion coefficient $D_{2d}$. Let us call the 118- the target and the 118+ the probe. We describe the target molecule as a disc of radius $R_1$ corresponding to half the end-to-end distance of the 118- chain. A given time *t*, $N_{target}(t)$ target molecules are embedded in the water-phenol interface, and the average area per target molecule can be estimated assuming that the distribution of target molecules on the interface of area *S* is homogeneous. The average area per target molecule is equal to:

$$A_{\text{target}}(t) = \frac{S}{N_{\text{target}}(t)} = \pi\left(R_2(t)\right)^2 \qquad 4$$

where $R_2(t)$ is the radius. The problem of the capture of a probe molecule by a target molecule corresponds therefore to a diffusion in a hollow cylinder of inner radius $R_1$ and outer radius $R_2$. The diffusion in a hollow cylinder has attracted considerable attention over time, and several solutions corresponding each to a particular boundary initial value have been proposed (*67, 71-73*). To our knowledge, these solutions deal only with situations where the radii $R_1$ and $R_2$ are independent of time. In these cases, the diffusion equation in cylindrical coordinates can be solved following the method of separation of space and time variables (*67*). In the present situation, when a probe encounters a target molecule, the formed product is expelled from the interface into the bulk. Therefore, this reaction is an interfacial annihilation reaction. As the reaction between the target and probe molecules proceeds, the water-phenol interface is depleted from target molecules and therefore the average area per target molecule increases. The present problem is that of the diffusion in a hollow cylinder with a moving boundary $R_2(t)$. Under these circumstances it is not possible to solve the diffusion equation by separating time and space variables. To overcome this mathematical problem we use a steady-state approximation expected to be valid at long times, when the concentration of target molecules becomes low enough. Under this assumption, the rate of capture of a probe molecule by a target molecule is such that the flux of probe molecules injected at $r$ (the in-plane polar coordinate) equals the flux $k_{\text{capture}}$ of probe molecules which disappear at $2R_1$:

$$k_{\text{capture}} = 8\pi R_1 D_{2d} \left.\frac{dP}{dr}\right|_{r=2R_1} \qquad 5$$

Where $P(r)$ is the probability to find a probe molecule at position $r$ (in plane polar coordinate) away from a target molecule. In a cylindrical symmetry the diffusion equation under steady state condition is written as:

$$\frac{d}{dr}\left(r\frac{dP(r)}{dr}\right) = 0,\ 2R_1 < r < R_2 \qquad 6$$

with boundary conditions $P(2R_1) = 0$, $P(R_2) = 1$. Such conditions imply that 1) the target surface concentration is very dilute and 2) that there is a reservoir of probe molecules.

Berg and Purcell (*74*) studied a similar problem for a plane with fixed boundary conditions. They concluded that the two-dimensional capture rate will dominate solely if the two dimensional rate of capture is much bigger than the three-dimensional rate. In our case the situation is somewhat different. Using the same approach as Berg and Purcell, we assume that the steady-state rate at the surface is reached when the rate of capture $k_{\text{capture}}$ at two dimensions is equal to the rate of adsorption at three dimensions. Using this assumption and the fact that the variation of the number of probe molecules $N_{\text{probe}}(z, t)$ in the $z$ (vertical) direction is independent of the in plane variations of the number $N_{\text{probe}}(r, t)$ of probe molecules, we can write an additional boundary condition to equation 6:

$$\nabla N_{\text{probe}}(r,z)\Big|_{z=0} = N_{\text{probe}}(r)\,\nabla N_{\text{probe}}(z)\Big|_{z=0} + N_{\text{probe}}(z=0)\,\nabla N_{\text{probe}}(r) = 0$$

Using the above considerations, the number of probes $N(r)$ for $2R_1 < r < R_2$ is therefore given by:

$$N_{\text{probe}}(r,t) = \left( \frac{dN_{\text{probe}}(t,z)}{dz}\Bigg|_{z=0} \right) \frac{\ln\left(\frac{r}{2R_1}\right)}{\ln\left(\frac{R_2}{2R_1}\right)} \qquad 7$$

where the first term in the bracket in the right hand side corresponds to the gradient of concentration of probes molecule in the direction normal to the interface. As the probability $P(r) = N_{\text{probe}}(r,t)/N_{\text{probe}}(r, t=0)$, the rate of capture is equal to

$$k_{\text{capture}} = \frac{\sin\left(\frac{\pi}{12}\right)\left(ah + \frac{h^2}{3}\right)}{V} \frac{32\pi^{\frac{3}{2}}}{\ln\left(\frac{R_2}{2R_1}\right)} \frac{D_{2d}}{\sqrt{D_{3d}}} \frac{1}{\sqrt{t}} \qquad 8$$

The rate of capture depends on time and on the number of target molecules. Two extreme cases can be considered. 1) If $R_2 = 2R_1$ the value of the capture rate diverges and goes to infinity. This condition corresponds to the close packing of target molecules at the interface. This situation corresponds to a very concentrated interface and is beyond the range of validity of the calculated rate of capture. However, the divergence of the two dimensional capture rate indicates that for a concentrated interface the two dimensional diffusion process is not the rate limiting step. 2) If $R_2$ is sufficiently larger than $R_1$, the situation corresponds to a dilute interface, where the steady-state approximation can be applied. Under this condition the two-dimensional rate of capture contributes to the overall rate of the reaction.

The variation of number of target molecule at the interface is equal to:

$$\frac{dN_{\text{target}}(t)}{dt} = -k_{\text{capture}}\left(t, N_{\text{target}}(t)\right)N_{\text{target}}(0) \qquad 9$$

where $N_{\text{target}}(0)$ is the initial number of target molecules at the interface. Combining equations 8 and 9, we obtain after integration:

$$\left[2\ln\left(\frac{R_s}{2R_1}\right) - \ln\left(N_{\text{target}}(t)\right) + 1\right]\frac{N_{\text{target}}(t)}{2} - \left[2\ln\left(\frac{R_s}{2R_1}\right) - \ln\left(N_{\text{target}}(0)\right) + 1\right]\frac{N_{\text{target}}(0)}{2}$$
$$= -N_{\text{target}}(0)\,64\pi^{\frac{3}{2}}\,\frac{\sin\left(\frac{\pi}{12}\right)\left(ah + \frac{h^2}{3}\right)}{V}\,\frac{D_{2d}}{\sqrt{D_{3d}}}\sqrt{t} \qquad 10$$

Where $R_s = \sqrt{\frac{S}{\pi}}$ is the radius of the interface. Under the close packing condition the number of target molecules lying on the interface is equal to the square of the ratio of the radius of the interface and the radius of the target ($N_{\text{close packing}} = \left(\frac{R_s}{R_1}\right)^2$). Under the hypothesis that the concentration of target molecule is very dilute at the interface ($N_{target}(t) \ll N_{close\ packing}$ at any time) equation 10 can be approximated to:

$$N_{\text{target}}(t) \approx -\frac{128\pi^{\frac{3}{2}}\,N_{\text{target}}(0)\sin\left(\frac{\pi}{12}\right)\left(ah + \frac{h^2}{3}\right)}{\left[2\ln\left(\frac{R_s}{2R_1}\right) + 1\right]}\,\frac{D_{2d}}{\sqrt{D_{3d}}}\,\frac{\sqrt{t}}{V} + N_{\text{target}}(0) \qquad 11$$

As pointed out above the recombination reaction at the interface can be considered as an annihilation reaction. Therefore the fraction of renatured ssDNA is equal to:

$$\theta_{\mathrm{dsDNA}}(t)=\frac{N_{\mathrm{target}}(0)-N_{\mathrm{target}}(t)}{N_{\mathrm{target}}(0)}=\frac{128\pi^{\frac{3}{2}}}{\left[2\ln\left(\frac{R_s}{2R_1}\right)+1\right]}\frac{\sin\left(\frac{\pi}{12}\right)\left(ah+\frac{h^2}{3}\right)}{V}\frac{D_{2d}}{\sqrt{D_{3d}}}\sqrt{t} \qquad 12$$

and therefore the fraction of renatured ssDNA should change as the square root of time.

As explained above the considered reaction has two kinetics components, the first one is the three-dimensional adsorption process and the second one is the two-dimensional process. These two processes correspond to the two asymptotic limits of the reaction (respectively at short and long times). In the intermediate regime the two kinetic components work cooperatively to form the dsDNA molecule. To model the data presented in figure 6.a, equation 12 should be modified. The two-dimensional steady-state can be considered to have been reached after a time $t_0$ where the $\theta_{\mathrm{dsDNA}}(t_0)$ fraction of dsDNA has already been formed. Under these conditions equation 12 should be modified, replacing $t$ by $t-t_0$ and adding a constant equal to $\theta_{\mathrm{dsDNA}}(t_0)$:

$$\theta_{\mathrm{dsDNA}}(t)=\frac{128\pi^{\frac{3}{2}}}{\left[2\ln\left(\frac{R_s}{2R_1}\right)+1\right]}\frac{\sin\left(\frac{\pi}{12}\right)\left(ah+\frac{h^2}{3}\right)}{V}\frac{D_{2d}}{\sqrt{D_{3d}}}\sqrt{(t-t_0)}+\theta_{\mathrm{dsDNA}}(t_0) \qquad 13$$

Equation 13 has been used to fit the last part of the experimental data presented on figure 8a. The value of the fitted parameters are presented below:

| Parameters | $\dfrac{D_{2d}}{\left[2\ln\left(\dfrac{R_s}{2R_1}\right)+1\right]}$ (cm$^2$ s$^{-1}$) | $t_0$ (s) | $\theta_{\text{dsDNA}}(t_0)$ |
|---|---|---|---|
| Values | $6.7 \times 10^{-10} \pm 1.4 \times 10^{-11}$ | $3390 \pm 60$ | $0.28 \pm 2 \times 10^{-3}$ |

Considering that the end-to-end distance of the target molecule is given by: $R_1 \approx \frac{1}{2} b N^{\nu}$, where $\nu$ is the Flory exponent ( $½ \leq \nu \leq 1$ in two dimensions) and $b$ is the size of a monomer ($b$ = 3.5 × $10^{-8}$ cm in our case), one can obtain lower and upper estimates for $R_1$ and from these bracket the two dimensional diffusion coefficient:

$3.2 \times 10^{-9} \pm 6.7 \times 10^{-11}$ cm$^2$ s$^{-1}$ $\leq D_{2d} \leq 3.8 \times 10^{-9} \pm 8.1 \times 10^{-11}$ cm$^2$ s$^{-1}$. The two-dimensional diffusion coefficient depends weakly (logarithmically) on the conformation of the ssDNA chains, leading to a narrow range of possible values: $3.5 \times 10^{-9} \pm 3.7 \times 10^{-10}$ cm$^2$ s$^{-1}$

*d. Renaturation with vortexing: PERT*

Figure 10a shows the kinetics of renaturation of the two complementary ssDNA 118+ and 118- (1.66 pM each) in a 100 µl mixture containing 90 µl of TE plus 0.85 M NaCl and 10 µl phenol vigorously shaken (2400 rpm). This corresponds precisely to the experimental conditions of PERT (*1*) (vigorous shaking of denatured DNA, equal amounts of complementary strands). The mixture was vortexed at 2400 rpm for various times, the reaction was quenched by the addition of enough TE buffer (100 µl TE buffer) to destroy the emulsion, and an aliquot was analyzed by gel electrophoresis. The reaction is quite fast: about two thirds of the DNA have renatured within 30 s. Nevertheless, by comparing the rates of adsorption and of renaturation (figure 7 and figure. 108.a), we see again that renaturation is the slower of the two processes. We know from the experiment on the rate of adsorption at this vortexing velocity that each complementary ssDNA is completely adsorbed within five seconds. In the renaturation experiment, all the data obtained after 5 seconds therefore involve complementary strands that have been previously completely adsorbed. The renaturation process between 5 and 30 s takes place exclusively at the water-phenol interface. This establishes the interfacial nature of PERT in this experiment. Assuming that the reaction follows a standard bimolecular kinetics, the reciprocal of $f_{ss}$, the fraction of DNA remaining single-stranded at time t is expected to grow linearly with t according to: $1/f_{ss} = k_2 C_0 t + 1$, where $k_2$ is the bimolecular rate constant, and $C_0$ the molar concentration of single-strands (here $C_0$ = 1.66 pM). The reciprocal of $f_{ss}$ has been plotted as a function of time in figure 10b. The rate constant obtained by a linear fit yields $k_2 = 4.2 \pm 0.4 \times 10^{10}\ M^{-1}s^{-1}$. This fit is only valid for times $t > 5\ s$. The exact meaning of this rate constant is complex: the solution is vigorously vortexed, and the three-dimensional diffusion process has become negligible. It is replaced by a convective flow that controls the arrival and the removal of material at the interface between water and phenol droplets. We calculated the rate of the reaction assuming a classical bimolecular process. The validity of this calculation can be questioned. The system is not under a steady state flow and the distribution of the reactants is not uniform in the solution. These inhomogeneities could lead to a hydrodynamic instability and in turn change radically the kinetics of the reaction (*75*).

# Discussion

We have studied an interfacial DNA renaturation, taking place at a water-phenol interface. We have used a decoupling scheme to establish the interfacial nature of this reaction. The decoupling scheme has led to a characterization of a salt-dependent interfacial confinement of single-stranded DNA, which was studied with and without shaking. We have tested 13 other organic solvents for their capacity of adsorbing ssDNA at the water-organic solvent interface at room temperature in the presence of NaCl. Phenol is the most efficient of all tested solvents. The study of DNA renaturation in the absence of shaking shows the existence of a regime where the rate of the reaction depends on the two-dimensional diffusion of the complementary single-strands. We have determined both the three-dimensional and two-dimensional diffusion coefficients of the ssDNA. The two-dimensional coefficient $D_{2d}$ is 10 times smaller than the three-dimensional coefficient $D_{3d}$. We have also investigated the effect of shaking on the interfacial renaturation. We established the interfacial nature of PERT in extremely dilute DNA solutions and measured the bimolecular rate constant for the reaction in these conditions. We will now discuss these results following the plan outlined in the introduction.

## 1) Adsorption and renaturation of DNA in the water-phenol two-phase system

*Water-phenol mixtures; salt effects on such mixtures.*

We study the behavior of nucleic acids in a water-phenol two-phase system, at various salt concentrations. The discussion of our data requires some knowledge of water-phenol mixtures with and without added salt. Mixtures of water and phenol have been avidly studied over the last century and a wealth of experimental data is available both for bulk ((*76, 77*) and further references therein, (*78*)) and interfacial properties (although only at the air-solution interface ((*79, 80*) and further references therein). At room temperature water and phenol are only partially miscible (9% phenol in water and 29% water in phenol in weight percent) (*78*). The two solvents become miscible above 66 °C (in other words the system has an upper critical solution temperature). The phase diagram has a complicated structure at low temperature (below 10 °C) (*77, 81*): it is closed by the formation of a phenol hydrate (with formula $(C_6H_5OH)_2H_2O$). The crystal structure of phenol hydrate has been determined (*82*). Phenol and water molecules are connected by hydrogen bonds, each water molecule being surrounded by four phenol molecules with a distorted tetrahedral structure.

Both adsorption and interfacial renaturation requires the presence of monovalent salts (typically in the order of one mole per liter for PERT (*1*)). The addition of salt modifies both bulk and interfacial properties of water-phenol mixture. Indeed, the addition of salt greatly enlarges the two-phase region in the temperature versus concentration water-phenol phase diagram (*76, 83-87*). The miscibility of the two solvents is decreased at fixed temperature, and the upper critical solution temperature is increased. The mutual solubility of water and phenol is decreased because the salt is preferentially soluble in water; this rule is sometime called Timmermans rule (*76*) (see (*88-90*) for theoretical modeling). At a fixed temperature the addition of salt lowers the solubility of phenol in the aqueous phase, from 9% without salt to about 5-6% in the presence of molar concentrations of salt (see for instance table 2 in (*1*)).

**Water-phenol nucleic acid mixtures**

How can we understand the interaction between phenol and nucleic acids in our system? Clearly, the presence of salt plays a key role, and the polyelectrolyte nature of the nucleic acids must be taken into account (*91, 92*). Also, we must distinguish between ssDNA and dsDNA. It is well known that phenol interacts better with ssDNA or single-stranded polynucleotides than with dsDNA: Leng *et al.* (*18*) studied the interaction between phenol and nucleic acids by mean of thermal stability, optical activity, viscosity and NMR measurements. They demonstrated that phenol interacts with denatured DNA, poly(A) and poly(U), but poorly or not at all with poly(C) and native DNA. The binding of phenol to single-stranded nucleic acids is therefore also sequence-dependent (we will not discuss this aspect in this work). The preferential binding of phenol to denatured DNA rather than to native DNA can also be inferred from studies of the DNA helix-coil transition in the presence of phenol: phenol lowers the melting temperature of dsDNA (*10, 11, 18, 93, 94*). Another argument is provided by our results: the dsDNA, once formed at the water-phenol interface is expelled in the aqueous phase. This shows that in contrast with the irreversibly adsorbed ssDNA, dsDNA does not interact efficiently with phenol. We will focus our discussion on the interaction of phenol with ssDNA. The behavior of nucleic acids in the presence of mixed solvents is often described using continuous approaches, (*95, 96*). We will examine the outcomes of continuous and discrete approaches for the understanding of phenol-ssDNA interactions, with or without salts. This will show the limitations of continuous approaches. We first briefly recall the polyelectrolyte properties of nucleic acids in aqueous solutions.

*Electrostatics of nucleic acids in aqueous solutions*

In neutral aqueous solutions each phosphate group of the backbone chain of dsDNA and ssDNA is dissociated and carries a negative charge. Nucleic acids are strongly negatively charged polyelectrolytes (see (*91, 92*) for reviews). In the absence of salt, the long-range electrostatic repulsion among the charged phosphate groups tends to destabilize the helical structure of the dilute dsDNA solutions (*91*) and to extend ssDNA. In the presence of moderate concentrations of added monovalent cations (typically between 10 mM to 1M), the range of the electrostatic interaction is screened, and the tertiary structure of dsDNA or ssDNA is altered. dsDNA has a local standard helical B structure (we neglect here sequence specific effects) and adopts a random or swollen coil conformation at large scale. ssDNA is much more flexible than dsDNA (*97*) and can have different types of local structure (helical or disordered). ssDNA can fold upon itself in a salt dependent manner, which results in intra-strand imperfect base pairing (*98-100*).

*SsDNA-Phenol interactions: continuous and discrete approaches*

The helical structure of ssDNA is the result of a balance between a destabilizing long range electrostatic repulsion and stabilizing short range stacking attractions. This qualitative description of ssDNA structure suggests that the configuration and the properties of ssDNA are mainly influenced by electrostatic and polarization forces. Two approaches can be chosen to describe these effects.

**a) Continuous approach**

*Without salt*

In this approach the milieu surrounding ssDNA is consider to be homogeneous and continuous. The strength of electrostatic and polarizing forces depends strongly on the ability of the solvent to be polarized. This solvent property is expressed through the dielectric constant of the medium (*101*). Water has a high dielectric constant ($\varepsilon_{water} = 80$) *i.e.* in an external electric field water molecules are easily oriented in the direction of the field. In the close neighborhood of ssDNA the strength of the electric field is important and therefore there exists a shell of oriented water molecules around the DNA molecule (*102-104*). The hydration shell will interact with the ssDNA's bases through hydrogen bonding and polarization interactions. Therefore the strength of stacking between bases should be affected by the quality of solvent-ssDNA interactions (*98, 105*). Phenol has a dielectric constant much lower than water ($\varepsilon_{phenol} = 10$ at 60° C (*106*)) and is partially miscible in water (9% in volume, at room temperature in the absence of salt). The effective dielectric constant of a phenol saturated water phase in the absence of salt can be estimated using the Clausius-Mossotti equation (also known in optics as the Lorentz-Lorenz equation (*101*)), and the values given above for $\varepsilon_{water}$ and $\varepsilon_{phenol}$. This yields an $\varepsilon_{solution}$ = 76.8 (without salt). Thus, the addition of phenol to the solution decreases the effective dielectric constant of the aqueous phase. According to Mel'nikov *et al.* this should lead to a weak compaction of the chain (*96*).

*With added salt*

The addition of salt into the solution decreases the amount of dissolved phenol in aqueous phase. At high salt concentrations (in particular at 0.85 M NaCl), we estimate that the percentage of phenol in the aqueous phase has dropped to 5%. This corresponds to an $\varepsilon_{solution}$ of about 77, close to the value of pure water. The increase of the dielectric permittivity induces a decrease of the local concentration of counterions surrounding ssDNA (*92*) and increases the Debye screening length. The addition of salt also decreases the range of electrostatic interactions. Under these circumstances, the destabilizing electrostatic repulsion should become less important than the stacking energy between bases that confers to ssDNA its local structure and its stability. Therefore one expects a folding of the ssDNA induced by the stacking the bases, which could involve imperfect intramolecular base-pairing as is observed in pure water (*98, 100*). To sum up, the continuous approach predicts a simple folding of the ssDNA molecule in the presence of phenol and salt, which would enhance intramolecular interactions (imperfect base-pairing).

**b) Discrete approach**

A discrete approach also takes into account forces at the molecular level (short range effect). We first briefly recall the role of short range interactions in the structure and stability of nucleic acids.

*Short-range interactions*

We will distinguish two types of short-range interactions: stacking and hydrogen bonding. DNA's bases are planar aromatic molecules and can form hydrogen bonds. Bases can interact together by their mutual polarization through a Π-Π interaction. This interaction is called Π-stacking. The competition between the short-range attractive stacking interaction between two adjacent bases and the long-range repulsive electrostatic interaction among the charged phosphate group of the backbone confers their helical structure to ssDNA and dsDNA. Watson-Crick hydrogen bonding between complementary bases (adenine with thymine and guanine with cytosine) further stabilizes the double-helical structure of dsDNA. The origin of these two short-range interactions (stacking and hydrogen bonding) lies in the capacity of the DNA bases to be polarized. The ability of DNA bases to create hydrogen bonds and the hydrophilic character of the anionic phosphate oxygen imply that water is an integral part of nucleic acid structure (*103*). The polymorphism of DNA (ssDNA or dsDNA) is intimately linked to the activity of water (*107*).

*Discrete approach without salt*

The above description treats the solvent as a continuous medium and does not take into account explicitly the interactions between the water, phenol and the polyelectrolyte. To describe the interaction of ssDNA with phenol molecule it is necessary to take into account its molecular nature. Water is a structured liquid (108). In the absence of phenol, in the close vicinity of ssDNA, the water molecule network follows the geometry of the polyelectrolyte (109, 110). The ability of phenol to form hydrogen bonds and its aromatic character allow this molecule to interact with water and to disturb locally the hydrogen network formed among water molecules. The interaction between water and phenol molecules can involve the hydrogen bonding observed in phenol hydrate crystals (82). In the presence of phenol, the water molecules in the vicinity of ssDNA must now both follow the geometry of ssDNA and interact with phenol molecules. This implies that phenol is in close vicinity of ssDNA. Therefore it is strongly favorable for phenol molecules to interact with ssDNA, either through a stacking interaction or hydrogen bonding. Phenol indeed increases the solubility of bases (13, 14): this shows the existence of a favorable interaction between phenol and the bases. On the other hand, dioxane and pyridine, two aromatic compounds with low dielectric constants have been shown by to decrease the solubility of pyrophosphate (10). This should also be the case for phenol: it should decrease the solubility of the phosphate backbone. As a result, the interaction of phenol with ssDNA is going to be a subtle balance of factors that either increase the solubility of the bases or decrease the solubility of the phosphate backbone.

The phenol molecule is repelled by the charged oxygen of the phosphate groups but can form a hydrogen bond with the oxygen linked to the phosphorus atom through a double bond. In the absence of salt, the stacking interaction between the bases in ssDNA is destabilized. This favors an intercalation of phenol between two adjacent bases. This could induce a change in the direction and the orientation of the bases with respect to the helical axis. In summary, in the absence of salt, phenol could form a complex with ssDNA involving a stacking interaction, occurring through an intercalation mechanism (111).

*Discrete approach with added salt*

The addition of monovalent counterions into the solution changes dramatically this picture. The concentration of dissolved phenol in the aqueous phase decreases as the concentration of added salt increases. SsDNA is a highly charged polyelectrolyte and the addition of cations into the solution screens the negative charge of the phosphate groups and induces the folding of ssDNA as discussed above. At low salt concentration (below the Manning's threshold (*92*)), the counterions are trapped along the phosphate backbone of the ssDNA. For ssDNA in the absence of phenol Manning's threshold (or local counterion concentration near the ssDNA) is equal to 0.21 M (*92*) (a similar value can be obtained by solving the Poisson-Boltzmann equation). Above Manning's threshold, electrostatic effects become less important. Both the repulsion between the charged phosphate backbone and the phenol molecule and the repulsion between two adjacent phosphates are decreased. Phenol molecules can more easily surround the ssDNA chain. The stacking energy between bases is on the order of a few $k_BT$ (Boltzmann's constant $k_B$ times the temperature $T$). In the absence of an electrostatic repulsion, bases can rotate freely around the phosphate backbone and can be stabilized by the phenol molecules present in the vicinity of the ssDNA chain. At this stage, we can only speculate on the conformation of this folded ssDNA: the helical structure of the DNA could be changed either by changing the direction of the helical turn or by having the phosphate backbone at the center with the bases completely exposed to the solvent and stacked with phenol molecules. In support of a stacking interaction, we can mention the results of Leng *et al.* (*18*). The authors have shown using proton magnetic resonance that phenol stacks with adenine bases of poly(A) and uracil bases of poly(U). Their experimental system and ours differ (they use a deuterated solution containing 0.11 M phenol and 0.25 sodium cacodylate), but their observation strongly favors a stacking mechanism in our system too. The presence or absence of an intercalating process cannot be inferred from their results. Another argument in favor of a stacking interaction between phenol and ssDNA comes from an analysis of the results obtained with the 13 other organic solvents (discussed below).

Figure 11 summarizes the two structures that we have discussed for a phenol-ssDNA complex: without salt or at low salt, the interaction would require an intercalation process, while at high salt (roughly above 0.2 M: see below), we expect stacking without intercalation. The existence of such complexes will have to be tested by spectroscopic of microscopic techniques. Remarkably, phenol has been shown to lower the viscosity of poly(A) in the presence of 1 M NaCl (*18*): this observation is not in favor of a simple intercalation mechanism, which should lead to an increase of the viscosity.

Similar structures have been proposed to describe protein-ssDNA in some filamentous bacteriophages (*112*), and have been observed in various single-stranded nucleic acid-protein complexes (*113*). These two ssDNA-phenol complexes correspond to different distances between two adjacent bases. The intercalated complex leads to a distance of about 7 Å, while in the second structure this distance could be still that found in B DNA.

The further increase in salt concentration (from 0.3 to 1 M) has two effects: 1) ssDNA folding, 2) expulsion of phenol from the aqueous phase. Both phenol and bases can interact with $Na^+$ through cation-π interactions (19). Phenol complexed through stacking to the bases becomes expelled from this phase, and this contributes to the transfer of the ssDNA from aqueous phase to the interfacial region between phenol and water.

**Conformation of the ssDNA in the water-phenol mixture**

*a) conformation in the aqueous phase: an interpretation of the three-dimensional diffusion coefficient of ssDNA*

Both fluorobenzene and phenol start to adsorb ssDNA at 0.3 M. This rules out a role for hydrogen bonding and suggests a simple electrostatic mechanism. According to this mechanism, the 0.3 M concentration corresponds to the local counterion concentration $C_{loc}$ near the ssDNA. Using Manning's equation (*92*), we calculate the distance $d$ between two adjacent bases:

$$d = \sqrt{\left[24.3\left(\frac{4\pi k_B T}{q^2 C_{loc}}\varepsilon_{solution}\right)\right]} \qquad 16$$

where $q$ is the protonic charge. This yields $d$ = 3.5 Å, close to the value found in B DNA. This is inconsistent with an intercalation mechanism (a distance of 7 Å would correspond to a concentration $C_{loc}$ = 0.08 M). This simple computation also supports a mechanism involving stacking without intercalation. Let us now try to understand the value of the diffusion coefficient in terms of a conformation of the ssDNA. We use a simple model where the ssDNA is approximated to a sphere of radius $R_G$, the radius of gyration, (to be more exact we should use the hydrodynamic radius of the ssDNA which is proportional to the radius of gyration;. however in a first approximation we consider both radii equal). Using the Stokes-Einstein relation, the diffusion coefficient is equal to:

$$D = \frac{k_B T}{6\pi\eta R_G} \qquad 17$$

where $\eta$ is the viscosity of the solvent. We consider that in the experimental condition the viscosity of the aqueous phase is the viscosity of water ($\eta = 0.98 \times 10^{-3}$ kg.m$^{-1}$.s$^{-1}$). From this equation we obtain a radius of gyration of 93 Å. This value can be related to the chain conformation using a simple scaling relation, valid for a flexible chain (*114*):

$$R_G \approx dN^{\nu} \qquad 18$$

where $d$ is the distance between two adjacent bases, $N$ the number of monomers (118) and $\nu$ is the Flory exponent ($1/3 \leq \nu \leq 1$) that defines the conformation of the chain. Combining equations 17 and 18, we derive a Flory's exponent $\nu \sim 0.68$. This value suggests that in the experimental conditions, the ssDNA chain has a swollen conformation. Our result is obviously crude and should only be viewed as indicative of the state of the chain.

*b) Conformation of ssDNA chains at the water-phenol interface.*

What is the conformation of ssDNA chains adsorbed at the water-phenol interface? Answering this question is important for understanding the mechanism of the interfacial renaturation. When a polymeric chain is adsorbed on a surface, it can retain its three-dimensional conformation if it becomes stuck on the surface. On the other hand, if it is able to rearrange freely on this surface, it can then spread and become a two-dimensional polymer. These two possibilities have been described for dsDNA spread on electron micrograph grids (*115*). Here, the chains are adsorbed at a liquid-liquid interface, on which they are able to diffuse, and the adsorption is irreversible. These three facts suggest that the adsorbed ssDNA chain at the water-are flattened and can be considered as two-dimensional polymers. One can envision the chains with their bases immersed in the phenolic phase, while the phosphate backbone and the sugars remain in the aqueous phase. There exist several possible simple conformations for the ssDNA chain: globular (swelling exponent $\nu = ½$), swollen coil ($\nu = 3/4$), or extended rod ($\nu = 1$). An adsorbed chain can be viewed as a disk of radius $R_1$ and occupies a surface $\pi(R_1)^2$. From this surface one can define an overlap concentration $C^*$ corresponding to the close packing of such disks. In the experiment where we renature DNA without shaking, renaturation becomes slower than adsorption when there remains about 0.47 ng of adsorbed 118- (corresponding to a release of about 20 % of the adsorbed material; see figure 8b). Now the onset of the decrease of the rate of renaturation should occur for a surface concentration close to the overlap concentration $C^*$. We can therefore use this experiment to guess the conformation of the adsorbed chain. For a globular conformation, ($R_1 = 0.5 \times 3.5 \times (118)^{1/2} = 19$ Å), we obtain $C^* = 580$ ng/cm$^2$, corresponding to an amount of chains of 58 ng on the interface. Assuming as swollen coil conformation ($R_1 = 0.5 \times 3.5 \times (118)^{3/4} = 62.6$ Å), we obtain $C^* = 50$ ng/cm$^2$, corresponding to an amount of chains of 5 ng. Finally, for an extended rod ($R_1 = 206$ Å), we obtain $C^* = 5$ ng/cm$^2$, corresponding to an amount of chains of 0.5 ng on the interface, very close to the value of 0.47 ng mentioned above. This simple reasoning suggests that the chains have an extended, rodlike structure at the interface. It is worth noting that a similar conclusion would still hold if we assume an intercalating mechanism (leading to a monomer size of 7 rather than 3. 5 Å). Indeed, with an intercalated chain having a swelling exponent ν ($R_1 = 0.5 \times 7 \times (118)^{\nu}$ Å), the close packing of 0.47 ng of ssDNA is obtained with $\nu \approx 0.85$ implying an extended structure.

*Comparison of phenol with other simple organic compounds: roles of hydrophobicity and aromaticity*

In addition to phenol, we have tested 13 organic solvents for their capacity of adsorbing ssDNA at the water-organic solvent interface at room temperature in the presence of NaCl. We will now compare our results with those reported in the literature, and discuss them in terms of hydrophobic and aromatic interactions. The interaction of simple organic compounds with DNA has been described by several authors (*10, 12, 17*). These authors have concluded that simple organic compounds interact with ssDNA mainly through a hydrophobic interaction, but the situation is actually more delicate.

According to Kauzmann (*9*), a hydrophobic interaction has a small and positive enthalpy variation and a large and positive entropy variation. In contrast, a stacking interaction between two planar aromatic compounds is not as a role entropically driven. Base stacking for instance is associated with a negative enthalpy and a negative entropy variation (*15, 116*). In other words, in water, it is more favorable for monomeric bases to form a stacked self-associated structure rather than to be excluded from water (*13*). This suggests that the description of ssDNA-organic compounds using a simple hydrophobic interaction must be carefully examined.

Levine, Gordon and Jencks (*10*) studied 54 organic compounds for their capacity to lower the melting temperature of dsDNA (destabilization of dsDNA in the presence of 0.04 M salt followed at 75°C). They came to the conclusion that the decrease of the melting temperature of dsDNA by these compounds is due to their preferential binding to ssDNA. The 6 most efficient compounds in their test were phenol, p-methoxyphenol, benzyl alcohol, aniline (four organic solvents) plus pyridine and purine. These six compounds are polar, planar, aromatic molecules. Levine, Gordon and Jencks concluded that the most efficient interaction with single-stranded nucleic acids is a hydrophobic interaction. They did not rule out a special mechanism for the action of aromatic compounds in their assay, but considered that this mechanism would be a minor contribution to an overall hydrophobic effect. In a similar study Ts'o and coworkers (*11*) examined 16 organic compounds for their capacity to lower the melting temperature of dsDNA or poly(A) and obtained similar result: their most efficient compounds were polar, planar, aromatic molecules such as purine, purine derivatives or phenol. Helmer, Kiehs and Hansch (*12*) analyzed the data of Levine *et al.* (*10*) by looking for correlations between octanol-water partition coefficients *P* of the compounds and their capacity to lower the melting temperature of dsDNA. They found that for the 12 alcohols and phenols as well as for the 5 amides used by Levine *et al.* there exist a good linear relationship between log (*C)* (where *C* is the molar concentration of the compound required to obtain 50% denaturation of dsDNA at 73°C in the work of Levine *et al.*) and log (*P)*. From these results they too concluded that these compounds act through hydrophobic interactions in the sense of Kauzmann.

The most efficient organic solvents in our adsorption assay are also all planar, aromatic, compounds. The three most efficient compounds are phenol, guaiacol (o-methoxyphenol) and fluorobenzene. Strikingly, phenol is the most efficient compound in these two very different assays. This underlines the efficiency of phenol as a ssDNA ligand. The most efficient compounds in the two assays are all planar, aromatic, polar molecules. The aromaticity of these compounds is therefore essential for an optimal interaction with single-stranded nucleic acids. Whether the interaction of ssDNA with these aromatic compounds is a typical (entropically driven) hydrophobic interaction will require further studies.

Aromaticity is of course not the only requirement. The presence of a hydroxyl group on the aromatic ring is also required, as witnessed by 1) the better efficiency of phenol over fluorobenzene at high salt concentrations and 2) the presence of two substituted phenols (o- and p- methoxyphenol) in the best performing compounds of the two assays. We conclude from these studies that the most efficient ssDNA binding compounds are planar, aromatic molecules and that phenol is the most efficient binding compound in these very different assays.

**Renaturation at the water-phenol interface**

We now discuss the mechanism of DNA renaturation at the water phenol interface.

*Topological issues in the interfacial renaturation*

The adsorption at the water/phenol interface is expected to lead to a reduction of dimensionality from to 3 to 2 for the ssDNA chain: this is the flattening discussed above. In that case, the topological constraints imposed by the reduced dimension of the interface could prohibit the formation of a complete helical structure of the dsDNA at the interfacial region. According to this hypothesis, when two complementary ssDNA encounter at the water-phenol interface, only a short (less than a full turn of the helix) double-stranded DNA segment is formed and at once transferred to the aqueous phase; the further growth of the double helix occurs at the same time as a transfer of the remaining ssDNA segments (see Figure 12.). The growth of the double helix in the aqueous phase is coupled with a two-dimensional translation and a rotation of the ssDNA segments. The adsorption of the extremities of ssDNA at the interface can prevent the topological entanglement of the two complementary chains during the annealing process. This could explain the efficiency of PERT in the renaturation of long DNA chains (even tens of kilobase long (*1, 3*)). On the other hand we note that even if the interface can accommodate a completely helical double stranded DNA, dsDNA is not expected to remain at this interface, since it partitions entirely in the aqueous phase in our system: the adsorption of dsDNA is not expected to be long lived.

*The interfacial mechanism of PERT*

As we have seen in the introduction, the basic mechanism of PERT, in particular the location of the nucleic acids in the two-phase system has remained poorly understood. According to Kohne et al. (1), "The single-stranded DNA may be concentrated at the phenol:aqueous interface or by forming aggregates or semiprecipitates elsewhere in the two-phase system". Other authors have favored the idea that the reassociation of nucleic acids takes place at the water-phenol interface, but were not able to exclude an aggregation/condensation mechanism (2, 3, 117). Our results obtained for extremely dilute ssDNA solutions establishes that PERT is an interfacial reaction. The complementary ssDNA are associated with phenol droplets when they renature. They are stably adsorbed at the surface of such droplets until recombination occur. We have shown that the interfacial adsorption is a monomolecular event in the absence of shaking. This result does not exclude the existence of aggregates in the aqueous phase either in the presence of shaking or at higher DNA concentrations.

*The role of shaking*

We have confirmed that shaking is necessary to obtain high reaction rates. The role of shaking is twofold. 1) It reduces the time required for the chains to reach the interface. 2) It greatly expands the interfacial area. We can give a rough estimate of this expanded interfacial area. We notice that the emulsion is turbid. This scattering of visible light implies that the size of the scatterer centers is in the micrometer range. A volume of 40 µl of phenol corresponds to about $10^{10}$ droplets with a radius in the micrometer range. This number is to be compared to the number of ssDNA chains (1.6 pM in 100 µl, about $10^{8}$ chains). The emerging picture is that there is a small portion of phenol droplets (about 1%) that carries a ssDNA chain. The annealing of complementary chains can only occur during the coalescence of two droplets.

As mentioned in the introduction, the rate of renaturation is independent of the length of the complementary strands at very low DNA concentrations (where the highest renaturation rates are obtained) (1, 2). The picture provided above for the mechanism of PERT in this case (where each chain is carried by an individual phenol droplet) suggests a simple explanation for this observation. The rate would be determined by the dynamic properties of these droplets and by the coalescence process rather than by the chains themselves.

*Hydrodynamics of PERT*

The PERT reaction involves the "vigorous shaking" of the water/phenol biphasic solution. The paragraph above implies that PERT depends in a crucial manner on the hydrodynamic field. The stability of the obtained microemulsion depends strongly on several parameters. The shape and the size of phenol droplets are defined by the combined action of the salt concentration and the rate of coalescence between phenol droplets. By considering the Laplace-Young theory of capillarity, the shape and the size of the phenol droplets can be linked to the interfacial tension between water and phenol and to the difference between the densities of the two bulk phases. The salt concentration influences not only the interfacial tension between water and phenol, but also the difference between their bulk densities. The diffuse ion layer (or Stern layer) that surrounds the phenol droplet depends on the salt concentration and could be one of the phenol microemulsion stabilizing factors. The coarsening rate of the mist of phenol droplets could be driven solely by salt concentration, if the mechanism of encounter between droplets were diffusion-controlled (*118*). However, because of the shaking, the flow is turbulent. The frequency of contacts between particles in such a flow increases substantially in comparison with the number of encounters in a motionless fluid. In addition, the frequency of encounters between droplets is not solely a function of the strength of shaking. For small enough particles (size of the particle smaller than the extent of the turbulent flow) there always exists a region where the Brownian diffusion predominates over the turbulent diffusion (*75*). Therefore, the frequency of encounters between phenol droplets will depend ultimately on both the strength and the frequency of the shaking, and also on the salt concentration, which determines the average size and shape of the droplets.

## 2) Comparison with other renaturating approaches

We will now compare the mechanism of this interfacial renaturation (with or without shaking) with other renaturation reactions. We distinguish three lines of investigation, which we describe now in detail.

*1) Thermal renaturation.*

The physical chemistry of nucleic acid annealing has been first studied in vitro in well-defined conditions using either synthetic polynucleotides (21, 22) or DNA (20, 23, 24). In all these works the reactions were investigated using dilute nucleic acid aqueous solutions in the presence of monovalent cations. This approach is called thermal renaturation. The rationale behind this choice was to try to avoid nucleic acid aggregation (119). As a result, thermal renaturation takes place in the bulk of a homogeneous aqueous phase. These classic investigations led to the conclusion that the reactions are thermally activated (hence their name), and proceed through a nucleation and growth mechanism (120). Monovalent salts have two effects; they increase the melting temperature of DNA (stabilize the dsDNA form) and increase the rate of renaturation (by lowering the electrostatic repulsion between complementary ssDNA). Salt concentration effects can be understood using appropriate laws of mass action (91, 92). These equations predict power law relations between salt concentration and the melting temperature or the renaturation rate, and such relations are observed experimentally (23, 91, 92). At elevated temperature, the rate of renaturation grows as the square root of the length of the complementary strands in the thermal renaturation of DNA (24). This is another example of a law of mass action behavior, where the square root exponent can be understood as an excluded volume effect reflecting the statistics of internal contacts in the ssDNA chains (121). At physiological temperatures (25-37°C), annealing in the presence of monovalent salt is inefficient for long ssDNA, because of imperfect intramolecular base- pairing (23, 98). One way to overcome this difficulty is to add organic solvents such as formamide or dimethyl sulfoxide in the solution (95, 117). In a variant form of thermal renaturation, one of the complementary strands is immobilized on a solid surface (117). The effect of this immobilization is to reduce the rate of the reaction compared to the bulk situation.

*2) Renaturation in the presence of condensing agents*

In a second line of investigation, a link between DNA renaturation and nucleic acid condensation was established (3). It has been shown that many simple compounds (incompatible polymers, polyamines, multivalent cations, cationic surfactants) that condense (i.e. collapse or aggregate) nucleic acids can also greatly accelerate DNA renaturation. As for monovalent salts, the addition of such condensing agents also increases the melting temperature of dsDNA (25, 122, 123). In the presence of the condensing agent, DNA renaturation takes place at the same time as the aggregation of the nucleic acids (single-stranded and double-stranded DNA) present in the solution (8). Renaturation of condensed nucleic acids can be kinetically first order or second order, depending on the nucleic acids concentration. Renaturation of extremely dilute nucleic acids condensed by spermine (8) can occur at rates compatible with a diffusion-controlled bimolecular reaction (118). The polyamines lower the electrostatic repulsion between the ssDNAs. It leads to large rate increases (25000 fold) at room temperature for small DNA chains (about 100 base pairs). The efficiency of the reaction decreases sharply for longer chains, because the intramolecular folding of the ssDNA greatly diminishes their accessibility (3, 25). In contrast with thermal renaturation, salt concentration effects on the rate of the reaction cannot be described using the law of mass action (3). Also in contrast with thermal renaturation, the rate of renaturation in the presence of multivalent cations has a weak temperature dependence (as expected for a reaction operating close to the diffusion control limit).

*3) Renaturation in the presence of SSBPs*

In a third line of investigation, single-stranded nucleic acid binding proteins (SSBP such as the gene 32 protein of bacteriophage T4 (*26*), RecA protein (*27*) or the *Escherichia coli* single-stranded binding protein (*Eco* SSB (*28*)) were purified and shown to be able to accelerate these reactions both under physiological temperatures and for long DNA. Many other single-stranded nucleic acid binding proteins have since shown to possess this activity. These proteins are often involved in genetic recombination (*30*), but also include various proteins such as the nucleocapsid protein of the human immunodeficiency virus I (*32*) and the prion protein (*34*). Single-stranded nucleic acid binding proteins facilitating nucleic acid annealing are often referred to as nucleic acid chaperones (*29*). Nucleic acid chaperones are generally essential to life (*33*).

The interactions of nucleic acid chaperones with nucleic acids and the mechanisms by which they accelerate their annealing are subtle, and can differ from one protein to another. Nevertheless, two general (non-exclusive) mechanisms have been described, which seem to account for some of the properties of several of these proteins. The first mechanism (A) involves the preferential interaction of the proteins with single-stranded nucleic acids, an interaction that greatly relies on aromatic amino acids; the second mechanism (B) involves the aggregation of single-stranded nucleic acids.

A) Nucleic acid chaperones bind in general preferentially to ssDNA rather than to dsDNA. Thus, in contrast with the monovalent salts or the simple condensing agents mentioned above, nucleic acid chaperones usually lower rather than increase the melting temperature of DNA.

Furthermore they stabilize single-stranded nucleic acids in an unfolded conformation. The removal of the imperfect secondary structure is presumably favorable for later annealing. Numerous experimental studies have shown that aromatic amino acids (phenylalanine, tryptophan and tyrosine) play an essential role in the interaction between SSBP and single-stranded nucleic acids (113, 124-127). The interactions between SSBPs and single-stranded nucleic acids fall in two categories. In the first category, the nucleic acids are immobilized on the protein surface. This can be observed in crystal structures of complexes between single-stranded nucleic acids and SSBPs, for instance for Eco SSB (113). The immobilization involves stacking interactions of aromatic residues on the nucleic acid bases. In the second category, the nucleic acid bound to the SSBP is still mobile. This has been seen for instance in the crystal structure of T4 gp32 complexed with an oligonucleotide (p(dT)6) which shows that the ssDNA is diffusing on the aromatic acid residues of the protein surface (125). Remarkably, depending on the experimental conditions, Eco SSB can form immobile or mobile complexes with ssDNA: mobile complexes have has been observed by NMR experiments (128).

B) Aggregation processes stimulate some of these proteins. It was proposed that the condensation of single-stranded nucleic acids also accounts (at least partially) for the acceleration of base pairing reactions by nucleic acid chaperones (3). Several nucleic acid chaperones have in fact been shown either to act as condensing agents, such the Escherichia coli RecA protein (5, 129), the protein NCP7 of HIV-1 (130, 131) or the prion protein (132), or to have an enhanced annealing activity in the presence of condensing agents (5, 28). These studies were usually performed at such DNA concentrations that the aggregation process was fast and the renaturation reaction kinetically first order.

*A) Comparison of PERT with the three lines of investigations: homogeneous and heterogeneous reactions*

Table 2 provides the values for the rate of annealing reactions of different nucleic acids in various conditions. The rate constants have been expressed using ssDNA chain molar concentrations (3). This allows a direct comparison of the constants obtained in the different experiments compared to the value expected for a diffusion-controlled process. Note that phosphate molar concentrations have usually been employed to express these rates (1, 23-26, 28, 117). The bimolecular rate constants span six orders of magnitude. The lowest rate constant, $4 \times 10^4$ $M^{-1}s^{-1}$, corresponds to the highest rate that can be obtained at room temperature in a dilute aqueous solution in the presence of monovalent cations for a long DNA molecule (T7 DNA, 40 kb long (23). In an aqueous solution, this rate can be greatly increased by raising both the temperature and the monovalent salt concentration (leading to a value of $1.2 \times 10^8$ $M^{-1}s^{-1}$at 65° C, 0,5 M NaCl). This large temperature dependence is the hallmark of thermal renaturation, and contrast with the weak temperature dependence of PERT. At the other extreme, the highest value is obtained using PERT with very dilute DNA solutions ((1), this work). The rate constants obtained with PERT in these conditions do not depend on the length of the DNA chains: we can therefore directly compare these rates with those obtained for T7 DNA at room temperature (23) We conclude that for long DNA chains, PERT can increase the rate of renaturation by a million-fold at room temperature. This rate acceleration is much larger than estimated earlier by Kohne et al. (an acceleration of "many thousand times" as mentioned in the introduction; this is because these authors compared the rate obtained with PERT with a rate obtained at an elevated temperature).

The comparison of these different rates in Table 2 requires care. The rate of PERT is measured under strong mixing conditions, whereas in the other experiments the samples are not stirred. For unstirred reactions, the theory of Smoluchowski (118) can be used to estimate the rate constant expected for a diffusion-controlled reaction, about $6 \times 10^9$ $M^{-1}$ $s^{-1}$. The rate constants obtained in the presence of various DNA condensing agents (polyethylene glycol (3), spermine (8), cetyltrimethyl ammonium bromide or CTAB (25)) are high enough to be compatible with a diffusion-controlled reaction in a dilute DNA solution. A diffusion-controlled base pairing reaction with modified bases in chloroform has been reported by Hammes and Park (133). Similar rates were obtained in the case of this simpler reaction.

The absolute value obtained for the rate constant using PERT is quite striking. Not only does PERT outperform the catalytic capacities of the single-stranded nucleic acid binding proteins gp32 and Eco SSB; the rate constant for PERT is also almost ten times faster than the Smoluchowski value predicted for a diffusion-controlled reaction. It is close to the value reported for the fastest (and much simpler) known bimolecular reaction, namely the recombination of $H^+$ with $OH^-$ ($1.3 \times 10^{11}$ $M^{-1}$ $s^{-1}$; (134)). This also shows that one can obtain rates that exceed a diffusion-controlled limit under appropriate mixing conditions. While such a conclusion is common knowledge in chemical engineering (135), we feel that it is not the case in molecular biology, where one often gets the impression that no process can be faster than a diffusion-controlled one. It is not generally appreciated, for example, that molecular motors can increase chemical rates beyond the diffusion-controlled limit (we plan to discuss this issue elsewhere). In PERT, the very fast rates require vigorous shaking. This shaking may be seen as a rudimentary version of the present day more sophisticated molecular motors.

In the presence of DNA condensing agents, the complementary ssDNA chains undergo an aggregating process at the same time as the renaturation reaction. We are not therefore dealing with a simple reaction that occurs in the bulk of a homogeneous aqueous solution, but rather with a reaction taking place at the same time as a phase separation. This remark points to a common feature between PERT and the reactions taking place in the presence of condensing agents. In both cases the reactions do not occur in the bulk of a homogeneous phase, in contrast with the standard (thermal) renaturation. In both cases also (adsorption using ¨PERT or aggregation with condensing agents), the volume available to the ssDNA chains is decreased and their effective concentration is increased, contributing to raise the rate of the reaction.

We therefore see that in Table 2, the concepts of homogenous chemistry can be used to account for the lowest rates (thermal DNA renaturation data of Studier (23)), whereas the highest rates come within the province of heterogeneous chemistry. The catalysis of renaturation by SSBPs corresponds to an intermediate situation in two aspects. 1) Homogeneous versus heterogeneous chemistry. Renaturation in the presence of condensing agents or the water-phenol two-phase system involves several macroscopic phases (DNA rich phases, phenol rich phase, phenol droplets of micron sizes). Renaturation in the presence of SSBPs involves the interaction of DNA with proteins of colloidal size.2) Renaturation rates. The catalysis by SSBPs leads to rate constants that are faster than those obtained in the presence of monovalent cations at "physiological" temperatures (25-35 °C), but smaller than those obtained with DNA condensing agents or with PERT. Note that the addition of the condensing agent spermine increases the efficiency of Eco SSB (increasing the rate constant by a factor 20 (28). This is likely to decrease the electrostatic repulsion between the ssDNA; whether this involves the condensation of the ssDNA chains is not known).

*B) Similarities between renaturation at the water phenol-interface and the renaturation in the presence of nucleic acid chaperones*

The mechanism of renaturation at the water-phenol interface (with or without shaking) contrasts with the standard thermal renaturation of nucleic acids in several respects such as temperature, chain length and salt concentration dependence. On the other hand, there exist striking similarities between this interfacial reaction and the mode of action of SSBPs:

1) Adsorption.

Both in this interfacial reaction and in the catalysis by SSBPs, the ssDNA chains must be adsorbed to a surface of the SSBP to react, (the surface of the protein or the water-phenol interface).

2) Preferential binding to ssDNA and stabilization of ssDNA

Phenol and SSBPs both bind preferentially to ssDNA. As a consequence, both phenol and (most) SSBPs lower the melting temperature of dsDNA. Aromatic residues of SSBPs play a key role in this preferential binding, as explained above. Remarkably, phenol stabilizes ssDNA at the water-phenol interface in an extended conformation, close to rodlike. This is reminiscent of the stiffening of ssDNA in the presence of RecA (136). A stabilization in an extended form is likely to contribute to the efficiency of the annealing reaction in the two systems.

3) Renaturation of long DNA chains

In contrast with simple condensing agents (3, 25), both PERT (1, 3) and SSBPs (see table 2) can fully renature very long DNA chains (tens of kilobase long).

3) Turnover

Both in the interfacial renaturation and in the renaturation in the presence of an SSBP, the formation of the double-stranded product of the reaction leads to a dissociation event: in the first case, the dsDNA is desorbed from the water-phenol interface, in the latter the SSBP dissociates from the dsDNA. This allows the reactions to turnover.

4) Surface diffusion of ssDNA.

The ssDNA chains can diffuse on the water-phenol interface. SssDNA oligonucleotides can also diffuse on the surface of different SSBPs (125, 128), Furthermore, in the case of gp32, this diffusion largely relies on tyrosine residues (125): thus both PERT and this protein make use of phenolic groups to provide the diffusive motion. In the case of Eco SSB, the interaction is more complex, involving other aromatic residues. In addition this protein can have various modes of binding, some of which lead to immobile ssDNA-protein complexes (31). We note that the surface diffusion of acetylcholine on the surface of acetylcholinesterase has been proposed to contribute to the high catalytic rates of this enzyme, and that this diffusion would also take place on a layer of aromatic residues (137). It has been proposed in this case that the aromatic side chains shield the direct electrostatic interactions between the substrate and the enzyme.

5) Dehydration of ssDNA.

Both in the water-phenol two phase system and in catalysis by SSBP, the ssDNA chains are dehydrated. Phenol stacking causes dehydration in the bulk of the aqueous phase, and the interfacial location further contributes to dehydrate the ssDNA. Extensive dehydration is also observed in complexes between double or single-stranded nucleic acids and proteins (138).

6) Salt concentration effects.

Both in PERT and in catalysis by SSBPs, increasing salt concentration does not lead to a simple law of mass action behavior, but to a more complex, cooperative effect. The existence of a coupling between adsorption and renaturation in PERT provides a plausible explanation for the observed critical dependence of the renaturation rate on the salt concentration (3). Indeed, the coil-helix transition should be roughened by a coupling with an adsorption process (35, 36, 139). This offers a plausible explanation for the observed breakdown of the law of mass action behavior. A similar reasoning (140) can be used to explain the salt effects in the case of a coupling between renaturation and a coil–globule.transition in the presence of condensing agents (3). In both cases the breakdown of the law of mass action is an indication of the heterogeneity of the system. For SSBPs, the situation is clearly much more complex (31). The results obtained with PERT suggest however that part of the cooperative effects seen with salts in the interactions between ssDNA and SSBP has its origin in the adsorption of the nucleic acids by these proteins.

*Comparison with other systems involving surface diffusion processes*

How does the diffusion of ssDNA at the water-phenol interface compare with other surface diffusion processes involving nucleic acids? A related issue is that of diffusion on surfaces and its possible role as a rate enhancing mechanism in biological processes (*74, 141, 142*). Adam and Delbrück have shown in a theoretical work that diffusion-controlled reaction rates between two reactants could be enhanced by a reduction of dimensionality of the reaction space. Several authors have extended this hypothesis to different biological and technological systems (*74, 143*). A typical scheme that Adam and Delbrück considered involves: *"Stage I, free diffusion in three dimensions to a specially designed surface (interface) in which the target is embedded. Stage II, diffusion on this surface. We imagine the molecule to be held on this surface by forces sufficiently strong to guarantee adsorption but also sufficiently weak to permit diffusion along the surface."* How is this scheme realized when the adsorbed molecule is a polymer? For a polymeric chain, even a small energy of adsorption per monomer will ultimately yield a large energy per chain. This is expected to flatten the chain and should greatly decrease the two-dimensional mobility ($D_{2d} << D_{3d}$). Another possibility is that the overall adsorption is weak enough to be reversible, but this would lead to a competition between surface diffusion and dissociation. It can be seen that the obtention of an irreversibly adsorbed chain with a decent surface diffusion is not an obvious task. In our case, the two-dimensional diffusion coefficient is only ten fold lower than the three-dimensional one. In contrast, in other systems the adsorption of nucleic acids can strongly reduces their surface mobility. An example is given by Guthold et al. (*144*), who have studied the diffusion of dsDNA restriction fragments deposited on mica. For a 1 kilobase pair fragment, they measure a diffusion coefficient $D_{2d}$ that is $10^{-6}$ smaller than $D_{3d}$ ($7 \times 10^{-14}$ $cm^2s^{-1}$ versus $5.4 \times 10^{-8}$ $cm^2s^{-1}$). The main difference between these experiments is that we deal with a liquid-liquid interface, whose fluctuations could enhance the interfacial mobility. This suggests that the fluidity of the interface is essential to obtain an efficient surface diffusion. This idea is supported by the experimental results reported by Maier and Rädler (*145*)Maier, 2000 #495]. These authors have studied the two-dimensional diffusion of dsDNA molecules on glass-supported cationic lipid membranes. For a 1 kilobase pair fragment, they measured a diffusion coefficient $D_{2d}$ of about $4 \times 10^{-10}$ $cm^2s^{-1}$ which is much larger than the $D_{2d}$ diffusion coefficient measured by Guthold et al. (*144*) for a DNA of the same size For an 80 base pair fragment, they measured a diffusion coefficient $D_{2d}$ of about $4 \times 10^{-9}$ $cm^2s^{-1}$ close to the value that we obtain for the 118- ssDNA chain ($3.7 \times 10^{-9}$ $cm^2s^{-1}$). We conclude that the scheme of Adam and Delbrück can be realized when the interface has a liquid-like structure. The interface can be a simple liquid-liquid interface, or the surface of a membrane. Another related issue is that of the diffusion of proteins of the surface of DNA (*146-149*). While this diffusion is commonly described as a one-dimensional process, it appears more realistic to consider it as two-dimensional one (because the DNA double helix has a finite width). It seems likely that the flexibility of both the DNA and the protein will contribute to the efficiency of the diffusion process. One way to test this hypothesis would be for instance to investigate the effect of the surface confinement of a DNA chain on the diffusion of the bound protein along it (*144*).

### 3) Implications for the partitioning of nucleic acids in water-phenol systems

The purification of nucleic acids in two-phase systems (water-chloroform, water-phenol or water two-phase systems) is one of the most common techniques of molecular biology (37-45). Most DNA extraction protocols have been designed for dsDNA (with the notable exception of the work involving ssDNA bacteriophages (41)). The goal of an extraction procedure is to provide intact nucleic acids, free of contaminants such as proteins or lipids. Phenol extraction is generally thought to be more efficient than chloroform extraction (47), probably because proteins partition completely in the phenolic phase (46, 150). However, phenol alone is often unable to disrupt complexes between nucleic acids and proteins, and added salts are required to obtain this disruption (151). This observation suggests that phenol is more a solvent than a denaturing agent for some proteins. Phenol alone is for instance often unable to inhibit RNase activity. Therefore, phenol is often used in combination with a potent protein denaturant (guanidinium thiocyanate) to obtain intact RNA (152, 153).

Our results have implications for the use of water-phenol extraction procedures in the isolation of DNA. Earlier studies reported losses of dsDNA (especially A-T rich DNA) due to denaturation and "aggregation" at the water-phenol interface (52-55). It was also shown that at low pH dsDNA is denatured and that the denatured ssDNA is transferred to the phenolic phase (46, 49). We have shown that a complete removal of single-stranded DNA material from the aqueous phase can occur because of its adsorption at the water-phenol interface. This adsorption does not require an aggregation process and does not necessitate a massive transfer of the nucleic acids in the phenolic phase, as is observed at lower pH. This suggests that in partitioning studies, a confusion has been made between an aggregation and an adsorption process. Our experiments show that a loss of ssDNA can occur at high pH, under conditions where dsDNA partitions entirely in the aqueous phase. We suspect that the original procedure used to isolate ϕX174 ssDNA (phenol extraction in the presence of a saturated sodium tetraborate solution, (41)) as well as other commonly used procedures lead to a significant loss of ssDNA through adsorption, and we will investigate this issue elsewhere. Presumably the loss of material only requires that a portion of the nucleic acids be single-stranded; in addition, the percentage of lost (adsorbed) DNA is likely to increase with the dilution of the DNA in the aqueous phase. These two conditions (presence of ssDNA portions in dsDNA, low concentrations) are encountered when working with degraded dsDNA, for instance in forensic research or when one isolates ancient DNA. We will mention as an example the extraction of DNA from Neandertal femur (performed in the presence of 0.5 M Sodium EDTA, pH 8, (154)). It is possible that the extraction procedures that are presently used in these fields lead to an overlooked loss of nucleic acids. We hope that our results will help design better extraction protocols for such situations.

## 4) Implications for the evolution of biological chemistry

The adsorption and renaturation of nucleic acids at interfaces is of interest for prebiotic chemistry for two reasons. First, a general problem of prebiotic chemistry is to explain the appearance of the first polymeric chains. Condensation polymerization offers a plausible mechanism. However bulk condensation polymerization reactions are known to be thermodynamically driven towards hydrolysis in dilute aqueous solutions, and also kinetically unfavorable if the concentration of monomers is low (61, 64). Thus, prebiotic chemistry is a field where the limitations of homogeneous chemistry have long been perceived. Two ideas borrowed from heterogeneous chemistry have been considered in great detail in prebiotic chemistry. The first one is Oparin's proposal of the role of coacervation. Coacervation is a liquid-liquid phase separation involving polymeric chains. Oparin favored the idea that prebiotic polymerization reactions took place in a heterogeneous, coacervated system, rather than in the bulk of a homogeneous phase (155). Another suggestion made by Bernal (60) and other researchers (62, 63, 156) is that prebiotic polymerization reactions took place on the surface of minerals such as clays. The adsorption of the monomers on the surface would both increase their local concentration and lower their bulk state of hydration. The adsorption of nucleotides and amino acids at the liquid-solid interface of clays has been shown to lead to their extensive polymerization, providing support for this adsorption scenario (62).

The second reason applies specifically to polynucleotides and to their replicating properties. Early ideas on the first replicating biopolymers in a primeval broth focused on polypeptides (155). Later one envisioned that the first self-replicating systems consisted of RNA and polypeptides (157). More recently, the discovery that RNA could be endowed with catalytic properties (158) has stimulated the idea of an initial era where the replicating systems consisted essentially of nucleic acids (57-59). Base pairing processes (both annealing and unwinding) are at the heart of these replicating systems, making helix-coil transition on surfaces processes of great interest. The interfacial chemistry of nucleic acids is therefore at the heart of prebiotic chemistry, because one can use it for both polymerization and replication reactions.

From our results and the analysis of the relevant literature, phenol emerges as an efficient compound in terms of its ability to interact single-stranded nucleic acids. We discuss now the implications of these findings for the evolution of biological catalysis.

*a) Prebiotic chemistry: a role for phenol in a nucleic acids world*

Here is a list of arguments in favor of a role phenol in a nucleic acid world.

1) Phenol is a plausible prebiotic compound

Phenol is known to be a plausible prebiotic compound: it can for instance be obtained from benzene, a compound present in interstellar matter (see (56) and further references therein).

2) Phenol binds efficiently to nucleic acids

The compounds active in a nucleic acids world are expected to be able to bind efficiently to nucleic acids. Phenol fulfills this requirement since it can interact with nucleic acids through aromatic interactions, not only in the bulk of an aqueous solution, but also at the interface between the water and the phenolic phases. Our results further show that phenol is the most efficient compound for this interfacial interaction among a variety of organic compounds.

We can apply the same reasoning on nucleic acid binding properties to the amino acids that can be obtained by prebiotic synthesis, which include the aromatic amino acids phenylalanine, tryptophan and tyrosine (see (159) for a review). We will mention here the results of a study of the interaction of immobilized amino acids with nucleotides (160). The aromatic amino acids phenylalanine, tryptophan and tyrosine were the most efficient compounds among nine amino acids tested in these experiments (this clearly favors a common role of a stacking interaction). Remarkably, tyrosine had the highest affinity for the five nucleotides 5'-Up, 5'-Cp, 5'-Ip, 5'-Ap and 5'-Gp (where U, C, I, A and G indicate uridine, cytidine, inosine, adenosine and guanosine (160): only tryptophan has a slightly better affinity for guanosine-5'-monophosphate). Thus, we conclude 1) that the remarkable affinity of the phenolic group for ssDNA reported here is also observed for the interaction of immobilized amino acids with nucleotides and 2) that the considerations on nucleic acids binding properties suggest an explanation for the later selection of aromatic amino acids among the twenty usual amino acids.

3) Adsorption on a liquid-liquid interface

Clays and other mineral surfaces have been shown to promote the polymerization of amino acids and nucleotides. In these materials, the catalytic surface is a liquid-solid interface. We have seen above that a lack of fluidity in this interface is poorly compatible with an efficient surface diffusion. Such surfaces also lack the flexibility which is found in present day enzymes, and which is known to be crucial to catalysis (161). This leads us to consider another type of primitive catalyst, where the catalytic surface is a liquid-liquid interface: the droplet of oil in an aqueous medium. Let us note at once that the likening of an enzyme with an oil droplet is not new: Perutz (162) has observed for instance that: *"the non-polar interior* *of enzymes provides the living cell with the equivalent of the organic solvents used by the chemists*". Our point of view is somewhat different because we emphasize the interfacial properties of the oil droplet. Furthermore the proposal that we want to make here concerns specifically the organic solvent that we consider, namely phenol.

4) Phenol droplets have a key catalytic activity

Phenol droplets have an annealing catalytic activity as shown by Kohne et al. (1) and here. Furthermore this annealing activity has been reported for both ribonucleic acids and DNA (1). It has been stressed that an important obstacle to the self-replication of nucleic acids is the presence of intramolecular stable structures (58, 163). Phenol could clearly help solve this problem. Thus, phenol has not only the expected binding properties but also a key required chaperone catalytic property. In addition, the release of dsDNA from the water-phenol after renaturation allows the reaction to turnover as mentioned before.

5) Phenol droplets can be generated in hydrothermal systems

Geochemical and microbiological evidence favor the hypothesis that early life has evolved near hydrothermal systems, at high temperatures (about 100 °C). This is the typical range of the critical solution temperature in a brine-phenol mixture (76). Heating and cooling cycles of brine-phenol mixtures above and below this temperature could have been used to produce large amounts of small phenolic droplets. Recently, Braun and Libchaber (164) have shown that thermal gradient can be used to increase the local concentration of DNA, and pointed out the relevance of this finding to the problem of prebiotic chemistry. There are interesting connections between the thermophoretic effect they describe and the mechanism that we propose for generating phenol droplets. In both cases, we obtain a large increase of the local concentration of nucleic acids, and this increase involves a convective effect rather than simple diffusion. The concentrating mechanism proposed by Braun and Libchaber (164) and the adsorption process envisioned here do not exclude one another, and could have operated simultaneously.

6) The chemistry of phenol could have been used in a nucleic acid world

The phenolic group is known to offer numerous possibilities in organic chemistry. The chemistry of phenol and of phenolic compounds could also have been exploited in a nucleic acid world. The ability of phenolic compounds to capture light energy suggests for instance that phenol or simple phenolic groups could have been used as early pigments, thus contributing to solve bioenergetic problems associated with prebiotic evolution (165). The linking of phenol to nucleic acids could also have been used to expand the catalytic properties of ribozymes: Robertson and Miller (56) have explored this possibility, and shown that a variety of nucleophiles, including phenol, can become bound to 5-hydroxymethyluracil under prebiotic conditions.

7) Phenol is ubiquitous in present day biochemistry

A last argument comes from the ubiquity of phenol and phenolic compounds (such as tyrosine) in present day biochemistry. Ubiquity would reflect phenol's antiquity. Several of the reactions involving phenol or phenolic groups today appear to be of interest from the point of view of prebiotic chemistry. The uibquinol/ubiquinone system for instance could be the relic of the early pigments mentioned above.

All these considerations lead us to propose that phenol droplets can catalyze the formation of polynucleotides under appropriate conditions; we hope that this hypothesis will soon be tested. While we believe that such interfacial catalysis could take place in a simple water-phenol system, we wish to point out that a simultaneous role of clays could also be tested. Indeed, clays are well known for their capacity to absorb phenolic compounds (this is a major theme in humus chemistry see (166) and further references therein), and it may also be the case that the presence of phenol increase the efficiency of clay-catalyzed polymerization.

*b) Early stages of life: a role for phenol in hyperthermophiles*

In the preceding section, we provided plausible arguments in favor of a role of phenol in prebiotic chemistry. More direct evidence can be offered at the later stage corresponding to the appearance of archaebacteria. Hyperthermophiles are considered to be closely related to the earliest forms of archeabacteria (*167*). The hyperthermophile *Ferroglobus placidus* can grow at 85 °C using phenol as the sole electron donor (*168*). Furthermore, the growth occurred in anaerobic conditions and using Fe(III) as the electron acceptor. These three features (use of simple aromatic compounds present in the deep, hot subsurface (*169, 170*), anaerobic growth and the use of Fe(III)) (*171*) are thought to have been important processes on the early Earth. Thus, these experiments provide a more direct support in favor of the role of phenol during the appearance of archaebacteria.

*c) Some remarks on the functions of phenol in present day biochemistry*

Our proposal that phenol played a key role in the origin of life may seem paradoxical if one recalls that phenol is well known as an antiseptic and a toxic compound. However, it must be stressed that the present day biochemistry need not bear a deep resemblance with prebiotic chemistry. The efficiency of phenol as an antiseptic varies with cell types: phenol is more toxic to eukaryotic cells such as leukocytes than to bacterial cells such as staphylococci or streptococci (*172*). The toxicity of phenol varies also greatly among bacterial cells (as discussed above, some extremophiles can grow on phenol at concentrations and temperatures that would be lethal to eubacteria). The same type of observations can be made concerning the ability of phenol to act as a denaturant. While many enzymes are denatured by phenol, there are reports of resistance to denaturation, for instance in the case of RNase, whose catalytic activity can withstand phenol extraction (*153*). Another example concerns the DNA polymerase of *Thermus thermophilus*, which retains its catalytic properties in the presence of several percent phenol at 55-65 °C (*173*)). A similar observation can be made on the peptidyl transferase activity of ribosomes (*174*). The catalytic activity is performed by a ribozyme (*175*). *Escherichia coli* ribosomes lose this activity upon phenol extraction; in contrast the ribosomes of the thermophilic eubacterium *Thermus aquaticus* retain 80% of this activity after treatment with proteinase K and SDS followed by vigorous extraction with phenol (*174*). All these experiments raise the possibility that the first catalysts (ribozymes or enzymes) had a resistance to the denaturing action of phenol that has been progressively lost in the course of evolution.

It is also important to note here that there are very few investigations dealing with the interactions of phenol with polypeptides. One of the reasons explaining the avoidance of phenol in denaturation studies is that it absorbs UV light and prevents the standard use of UV spectroscopy to monitor unfolding. The comprehensive review of Singer (*176*) on the properties of proteins in nonaqueous solvents makes no mention of experiments involving phenol. More recently, Zaks and Klibanov have performed striking experiments, showing that a number of enzymes are active in a variety of organic solvents (*177, 178*). Phenol was again not among the solvents tested. Clearly, there is room for additional work in this field.

## Summary and concluding remarks

*Summary of the experimental results*

We examined the room temperature behavior of very dilute solutions of two complementary 118 base-long single-stranded DNA chains in a two-phase system containing water, phenol and sodium chloride. 1) In the absence of its complementary strand, each single-stranded DNA chain can be adsorbed to the water-phenol interface in a salt-dependent manner. At 0.85 M NaCl, 80% of the chains are irreversibly adsorbed. In the absence of shaking, this adsorption is a slow, monomolecular, diffusion-controlled process. The adsorption is completed within a few seconds if the solution is vigorously shaken. In addition to phenol, we tested thirteen other organic solvents for their ability to adsorb single-stranded DNA at the water-solvent interface. Phenol is the most efficient solvent, followed by other planar aromatic compounds. This supports a role for phenol stacking in its interaction with single-stranded DNA. 2) The addition of a single-strand in a brine-phenol two-phase system (0.85 M NaCl) where the complementary single-strand has been previously adsorbed initiates an interfacial renaturation reaction, coupled to the release of the renatured double-stranded DNA in the aqueous phase. In the absence of shaking, the adsorption of this added single strand and DNA renaturation occur at similar rates at short time; at longer time the two-dimensional diffusion of the complementary chains governs their reassociation. 3) The addition of the two complementary chains in the aqueous phase of this brine-phenol system followed by a constant, vigorous shaking leads to a fast interfacial renaturation. We have determined a bimolecular rate of renaturation equal to $4.2 \pm 0.4 \times 10^{10}$ $M^{-1}s^{-1}$; this rate greatly exceeds the Smoluchowski limit for a three-dimensional diffusion-controlled reaction.

*Summary of the discussion*

DNA renaturation in a vigorously shaken water-phenol two-phase system is known as the Phenol Emulsion Reassociation Technique or PERT (1). Our results show that for very low DNA concentrations, PERT is an interfacial reaction. The existence of a coupling between adsorption and renaturation accounts for the unusual salt dependence of the reaction. A comparison of PERT with other approaches used to obtain high renaturation rates (involving either simple DNA condensing agents or single-stranded nucleic acid binding proteins) shows that PERT is the most efficient technique and reveals similarities between PERT and the renaturation reaction performed by single-stranded nucleic acid binding proteins. The most efficient renaturation reactions (in the presence of condensing agents or with PERT) do not occur in the bulk of a homogeneous phase. Rather, they belong to the field of heterogeneous chemistry. Our results lead to a better understanding of the partitioning of nucleic acids in two-phase systems, and should help design improved extraction procedures for damaged nucleic acids. We discussed the role of phenol and liquid-liquid interface from the point of view of prebiotic chemistry and today biochemistry. We have seen that fluid interfaces can allow efficient surface diffusion processes. We reached the conclusion that phenol and phenol droplets could have played a central role both in prebiotic chemistry and in the early stages of life.

*Concluding remarks*

*Homogeneous and heterogeneous approaches to nucleic acid biochemistry.*

The current description of nucleic acid biochemistry is mainly based on *in vitro* experiments performed in the bulk of homogeneous, dilute aqueous solutions (*91, 179-183*). In chemistry, the investigation of chemical reactions distinguishes between homogeneous chemistry, where the reaction proceeds in the bulk of a homogeneous phase, and heterogeneous chemistry, where the reaction involves more than one phase. In the same manner, one can draw a distinction between homogeneous biochemistry and heterogeneous biochemistry. Clearly, the current description of nucleic acid biochemistry is mainly based on homogeneous biochemistry. In contrast, we have shown here that the most efficient renaturation reactions pertain to heterogeneous biochemistry. Furthermore, our study of the water-phenol interfacial renaturation reveals striking similarities with the catalysis by SSBPs. This implies that homogenous approaches in nucleic acid biochemistry have limitations, and that the study of heterogeneous system should lead to a better understanding of the behavior of nucleic acids, both *in vitro* and *in vivo*. We plan to examine this issue elsewhere.

**Acknowledgments**

We are grateful to C. Mann, O. Hyrien, T. Odijk and Y. Timsit for discussions and for their help in improving the manuscript. We also thank A. Braslau, M. Daoud, P. Guenoun, M. Olvera de la Cruz, E. Yeramian and G.Zalcser for useful discussions.

| Category | Compound | Critical salt concentration for ssDNA adsorption | SsDNA in aqueous phase at 3 M NaCl (%) |
|---|---|---|---|
| HYDROPHOBIC | Dodecane | - | 100 |
| | Carbon tetrachloride | - | 100 |
| | Dichloromethan | 2.5 | 51 |
| | Chloroform | 2.5 | 33 |
| AROMATIC | Toluene | - | 100 |
| | Aniline | - | 100 |
| | Bromobenzene | 1.4 | 41 |
| | Iodobenzene | 1.4 | 54 |
| | Chlorobenzene | 1 | 57 |
| | Fluorobenzene | 0.3 | 22 (2.5 M) |
| | Guaiacol | 0.5 | 10 |
| | Phenol | 0.3 | 0 |
| PRIMARY ALCOHOL | Butanol | - | 100 |
| | Benzylalcohol | 2 | 48 |

**Table 1.** Partitioning of 118- for 14 organic solvents. The salt concentration at the onset of the adsorption was determined, as well as the percentage of 118- remaining in the aqueous phase at 3 M NaCl (except for fluorobenzene, where it was determined at 2.5 M NaCl). The organic compounds are sorted by increasing efficiency in each sub-group.

| Nucleic Acid | Reaction medium | Temperature | $k_2$ ($M^{-1}s^{-1}$) | Reference |
|---|---|---|---|---|
| Intact Phage T7 DNA (39.9 kb) | 0.1 mM buffer, pH 8<br>40 mM NaCl | 25 °C | $4 \times 10^4$ | Studier (1969) |
| Intact Phage T7 DNA (39.9 kb) | 0.1 mM buffer, pH 8<br>60 mM NaCl | 35 °C | $6 \times 10^5$ | Studier (1969) |
| Intact Phage T7 DNA (39.9 kb) | 0.1 mM buffer, pH 8<br>0.5 M NaCl | 65 °C | $1.2 \times 10^8$ | Studier (1969) |
| Phage T4 DNA (168.9 kb) ~ 16 kb long fragments | 20 mM buffer, pH 7.6<br>10 mM KCl<br>40 mM $MgSO_4$<br>gp32 (in excess over DNA) | 37 °C | $1 \times 10^8$ | Alberts and Frey (1970) |
| Intact Phage λ DNA (48.5 kb) | 10 mM buffer, pH 5.5<br>10 mM NaCl<br>10 mM $MgCl_2$<br>*Eco* SSB (1.2-fold excess over DNA) | 37 °C | $1.6 \times 10^7$ | Christiansen and Baldwin (1977) |
| Intact Phage λ DNA (48.5 kb) | 10 mM buffer, pH 5.2<br>20 mM NaCl<br>*Eco* SSB (10-fold excess over DNA)<br>2 mM spermine | 37 °C | $3.4 \times 10^8$ | Christiansen and Baldwin (1977) |
| 118 bp DNA fragment from phage *ϕ*X 174 | 10 mM Tris HCl<br>1 mM EDTA<br>0.46 mM spermine | 25 °C | $1.4 \times 10^9$ | Chaperon and Sikorav (1998) |
| 124 bp DNA fragment from pSV2gpt | 50 mM NaCl<br>1 mM CTAB | 68 °C | $6 \times 10^9$ | Pontius and Berg (1991) |
| *E. coli* DNA 0.5 µg / ml | PERT, 2 M LiSCN | 22-24 °C | $4.2 \times 10^{10}$ | Kohne *et al.* 1977 () |
| 118 bp DNA fragment from phage *ϕ*X 174 | PERT, 0.85 M NaCl | 22-24 °C | $4.2 \times 10^{10}$ | This work |

**Table 2.** Bimolecular rate constants of DNA renaturation in various experimental conditions.

## Figure legends

**Figure 1.** Partitioning of 118- as a function of NaCl concentration: aqueous phase (○); phenolic phase (Δ); adsorbed on the wall of the tube (□); water-phenol interface (■).

**Figure 2.** Left. Picture of a 15 ml silanized tube containing a mixture of brine (0.85M NaCl in TE buffer solution) and phenol (40% in volume) plus 0.125 nM $^{32}$P labelled 118-. Top arrow: air-water interface; bottom arrow: water-phenol interface. Distance between the two arrows: 6 cm. Right. Autoradiography of a portion of the tube, showing a peak at the water-phenol interface (arrow).

**Figure 3.** Reversibility of the adsorption of 118+ as a function of NaCl concentration. Percentage of 118+ in the aqueous phase: adsorption (○); desorption (■).

**Figure 4.** Percentage of 118- remaining in the aqueous phase as a function of salt concentration for three solvents: phenol (○), guaiacol (□) and fluorobenzene (Δ).

**Figure 5.** a) Release of adsorbed 118- by the complementary polynucleotide 118+. Percentage of released 118- as a function of added 118+ (○); Control experiment: percentage of released 118- as a function of added 118- (Δ). b) Initial percentage of released 118- as a function of added 118+: experimental data (○), linear fit (---- ----).

**Figure 6.** a) Percentage of 118- (0.125 nM) removed from the aqueous phase in the absence of shaking in a biphasic mixture of brine (0.85M NaCl in TE) and phenol (40% in volume) as a function of time. Experimental data (○). The dashed line is the fit obtained with Equation 3.

Inset: Initial percentage plotted as a function of square root of time. b) Schematic drawing of the experimental tube, $\boldsymbol{a}$ = 0.5 cm and $\boldsymbol{h}$ = 0.8 cm.

**Figure 7.** Adsorption by vortexing of 118- (0.125 nM) at the brine (0.85M NaCl in TE)-phenol (40% in volume) interface as a function of time of vortexing for various vortexing velocities (in rounds per minute).

**Figure 8.** a) DNA renaturation without shaking of the two complementary polynucleotides 118+ and 118- (0.125 nM each) at the brine (0.85M NaCl in TE)-phenol (40% in volume) interface as a function of time. Experimental data (○). The dashed line (----) is the fit obtained with Equation 13. b) Comparison of the rate of DNA renaturation without shaking (data of Figure 6.a) (■) and of the rate of adsorption of 118- at the brine-phenol interface (data from Figure 4.a) (Δ) as a function of square root of time. The two curves have been rescaled in order to be superimposed at short times

**Figure 9.** The two kinetics paths **A** and **B** of the renaturation reaction at water-phenol interface in the absence of vortexing. **A**) 1 The probe molecule (**—**) encounters directly the target molecule (–) through a 3-dimensional diffusion, 2- The dsDNA is formed. 3- The dsDNA form is expelled from the interface to the aqueous phase. **B**) 1'- The probe molecule is first adsorbed at the interface through a 3-dimensional diffusion. 2'- The probe molecule and the target molecule encounter each other at the interface through a 2-dimensional diffusion. 2"- The dsDNA is formed. 3- The dsDNA form is expelled from the interface to the aqueous phase.

**Figure 10.** DNA renaturation by vortexing of the two complementary polynucleotides 118+ and 118- (0.125 nM each) at the brine (0.85M NaCl in TE)-phenol (40% in volume) interface. a) Percentage of renatured DNA as a function of time. b) Time variation of the inverse of the survival probability (1/fss) showing the experimental data (○) and the linear fit.

**Figure 11.** Schematic drawing of the possible structures of ssDNA in presence of phenol molecules. a) ssDNA in the absence of phenol molecules. Bases are stacked together and the distance between two adjacent bases is 3.4 Å. b) A hypothetic intercalated structure of ssDNA in the presence of phenol molecules (distance between two adjacent bases is 7 Å). c) A hypothetic stacked structure, without intercalation, of ssDNA in the presence of phenol molecules.

**Figure 12.** The hypothetic transient structure formed during the interfacial renaturation. As two regions of two complementary ssDNA encounter each other a double helix is formed. The double helical portion (assumed here to have started inside the ssDNAS rather than at one extremity) is progressively expelled from the water-phenol interface to the aqueous phase. The single-stranded portions still remain adsorbed at the interface.

Figure 1

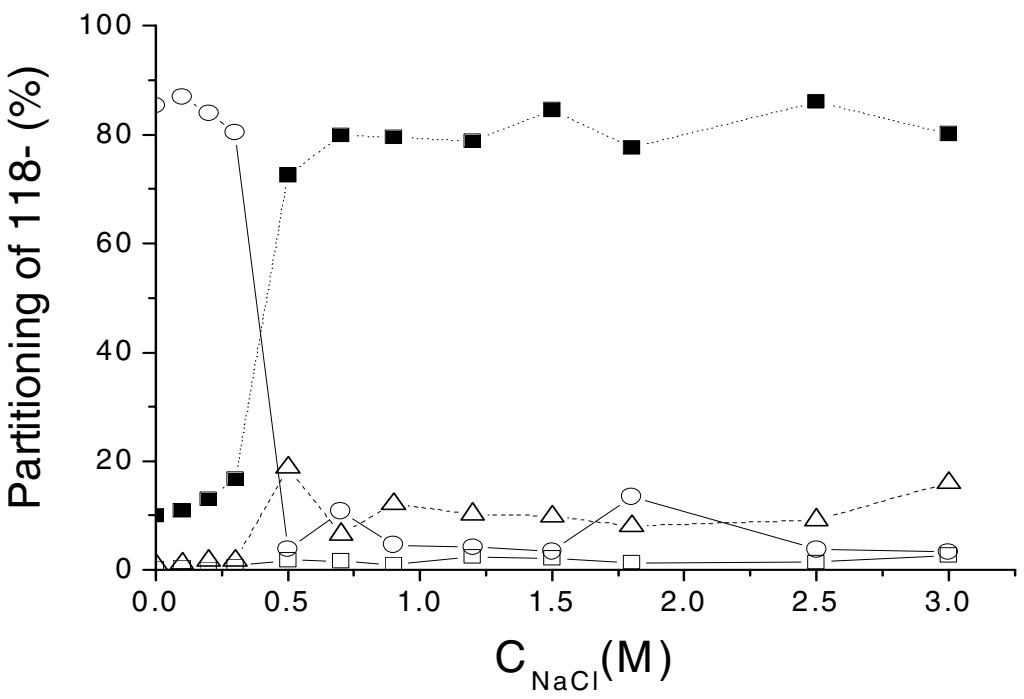

Figure 2

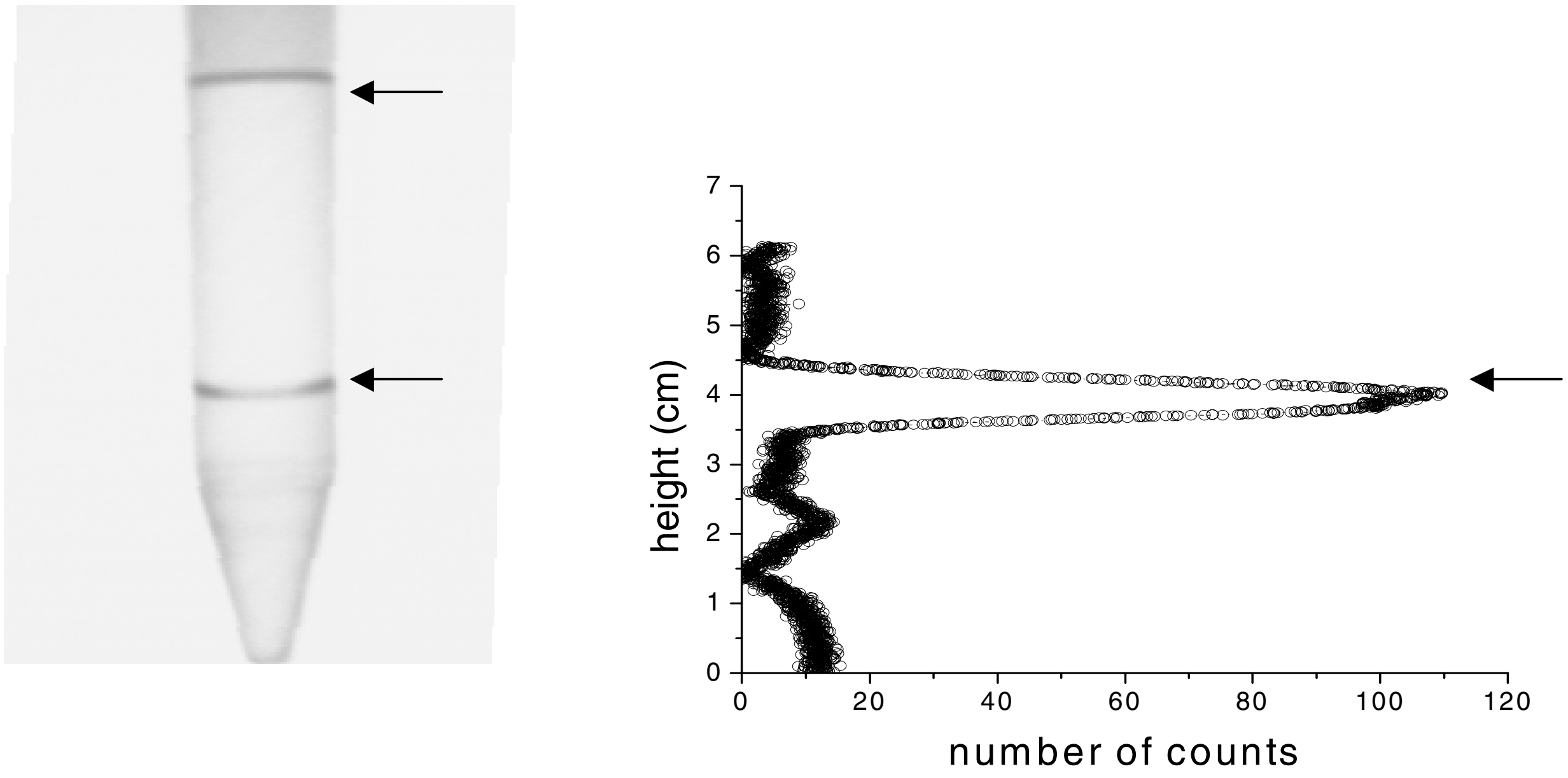

Figure 3

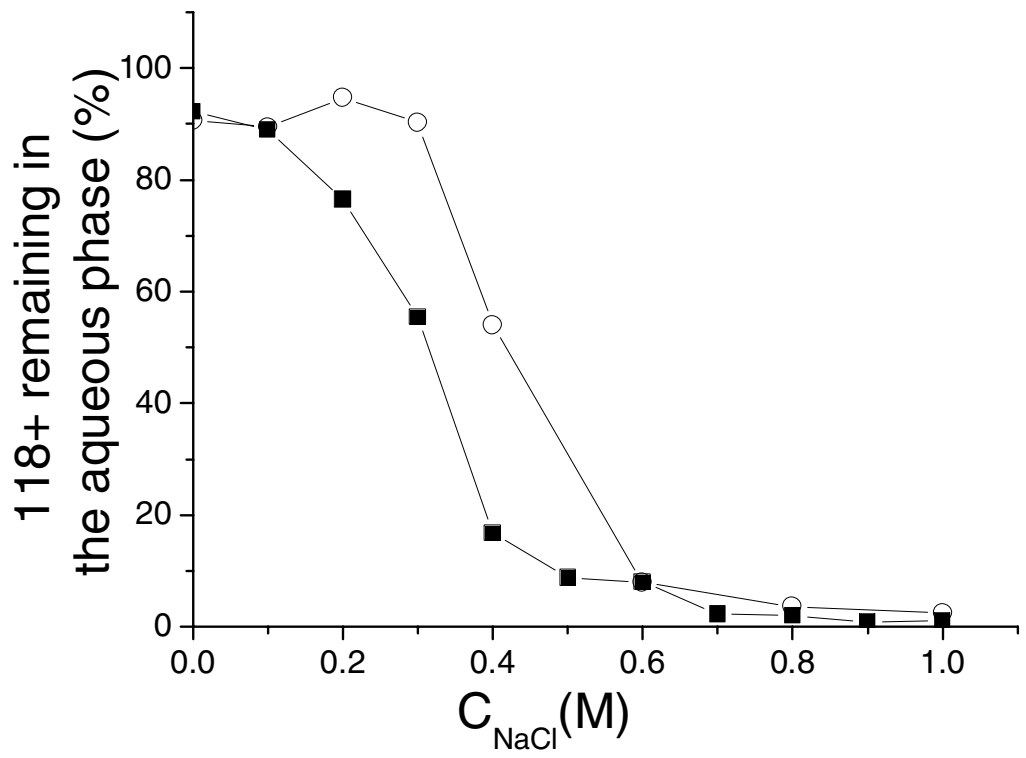

Figure 4

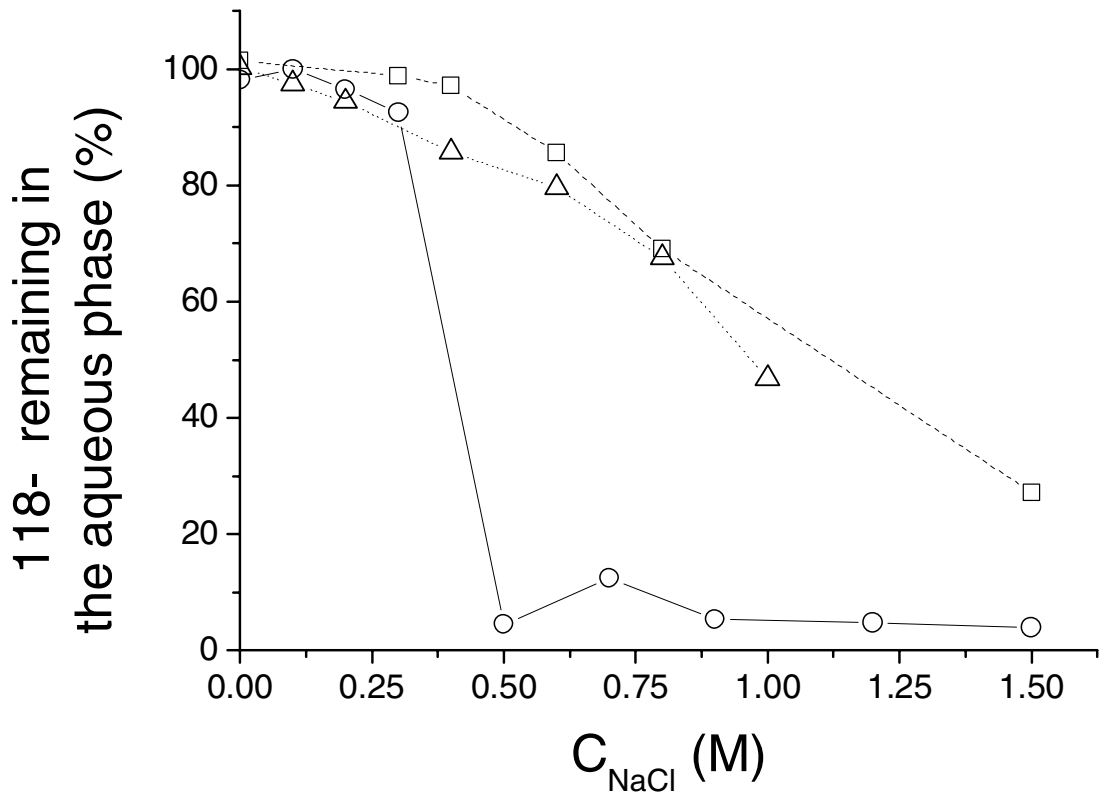

Figure 5a

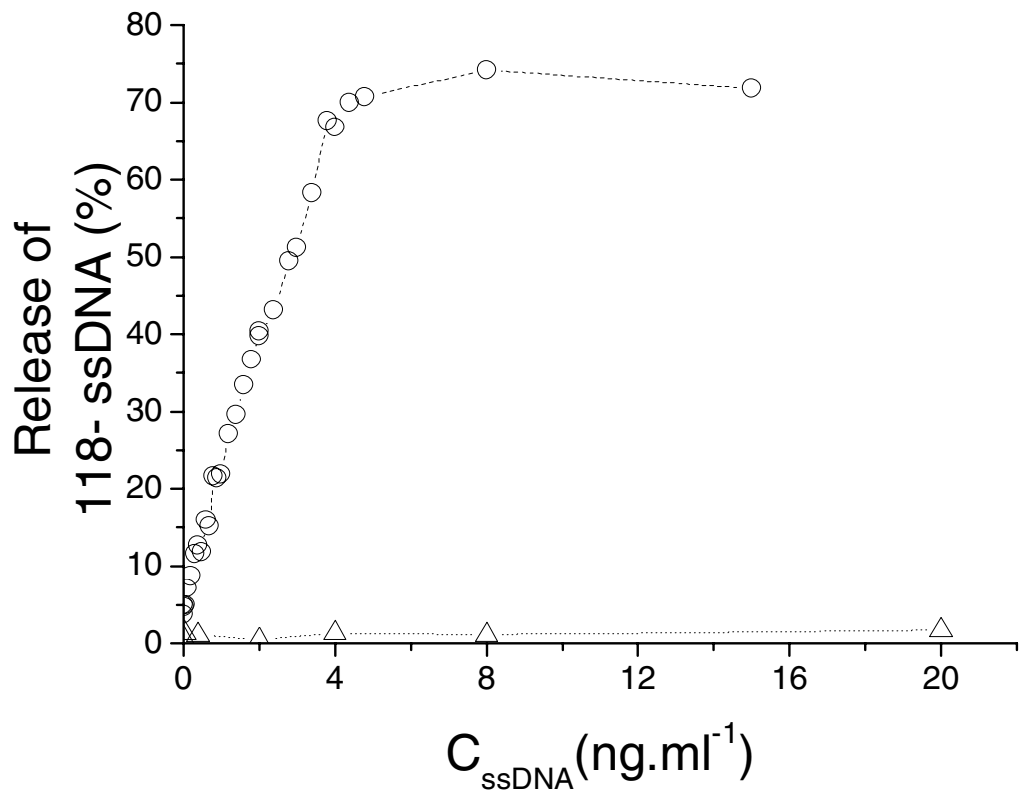

Figure 5b

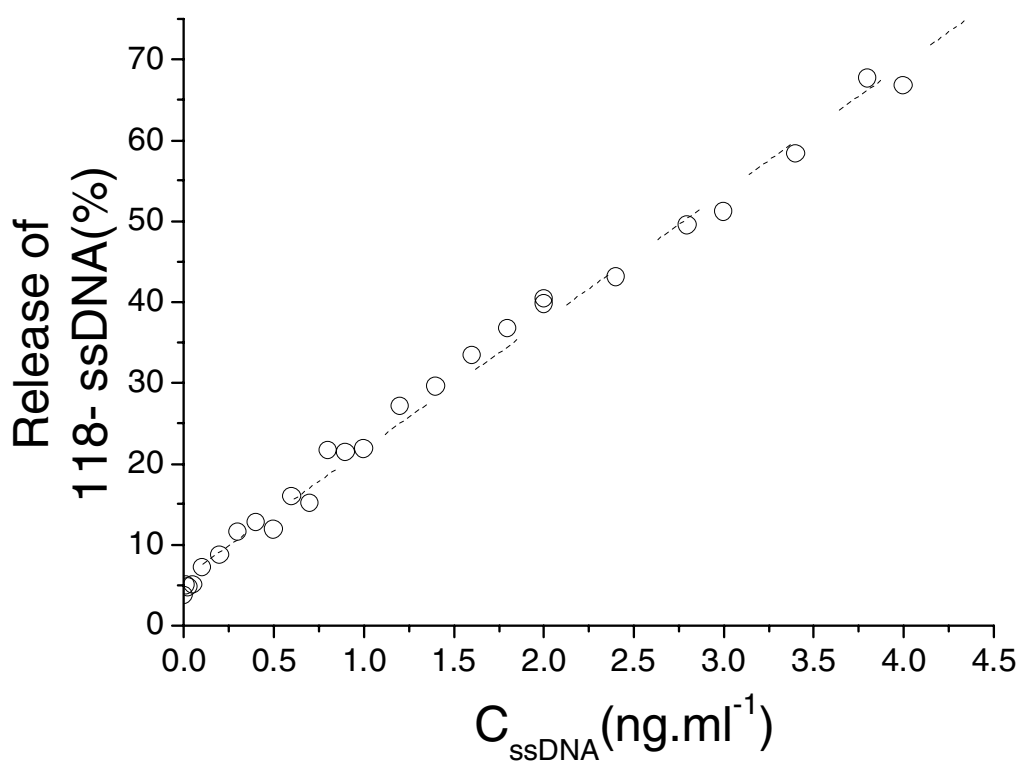

Figure 6a

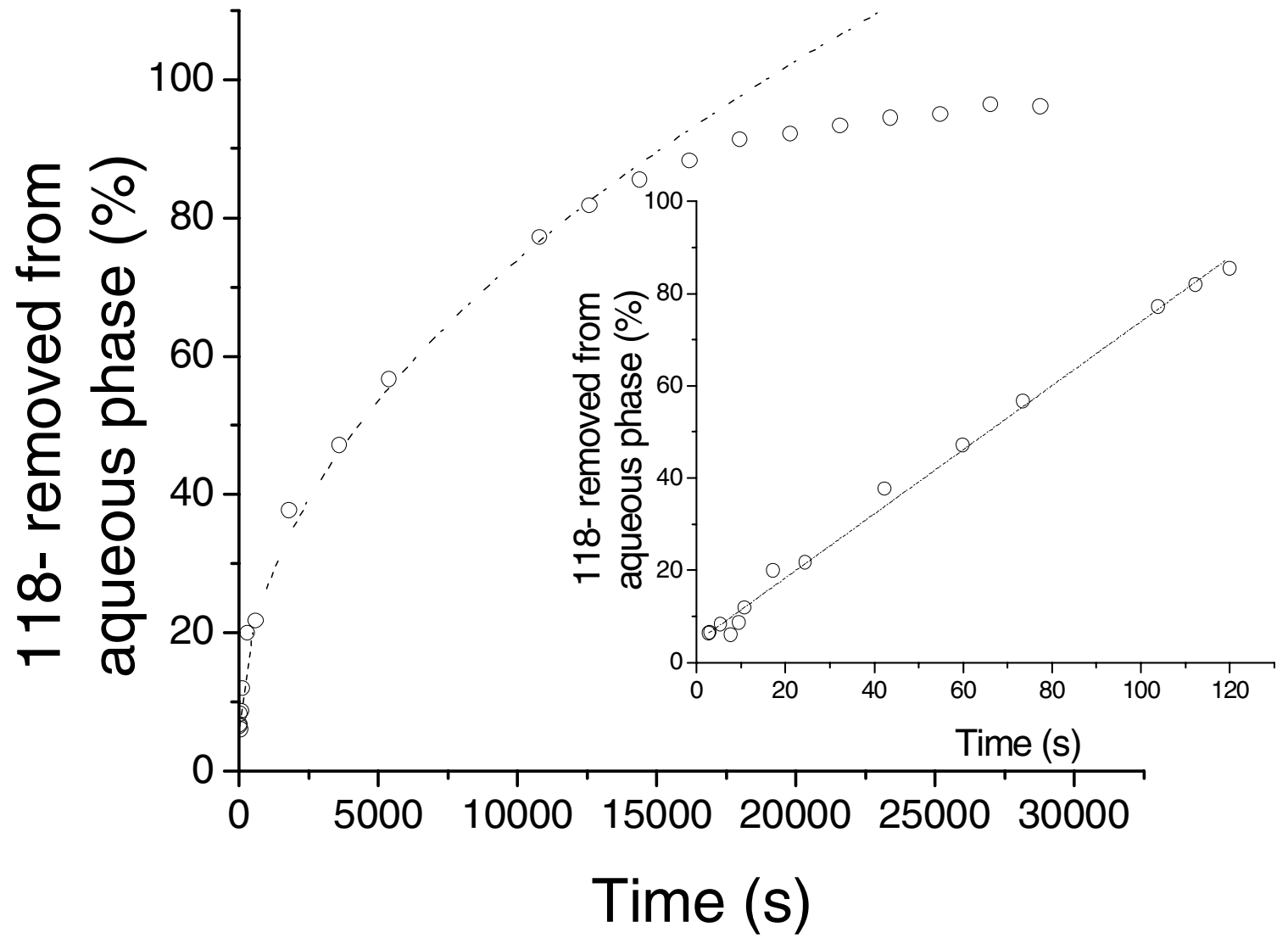

Figure 6b

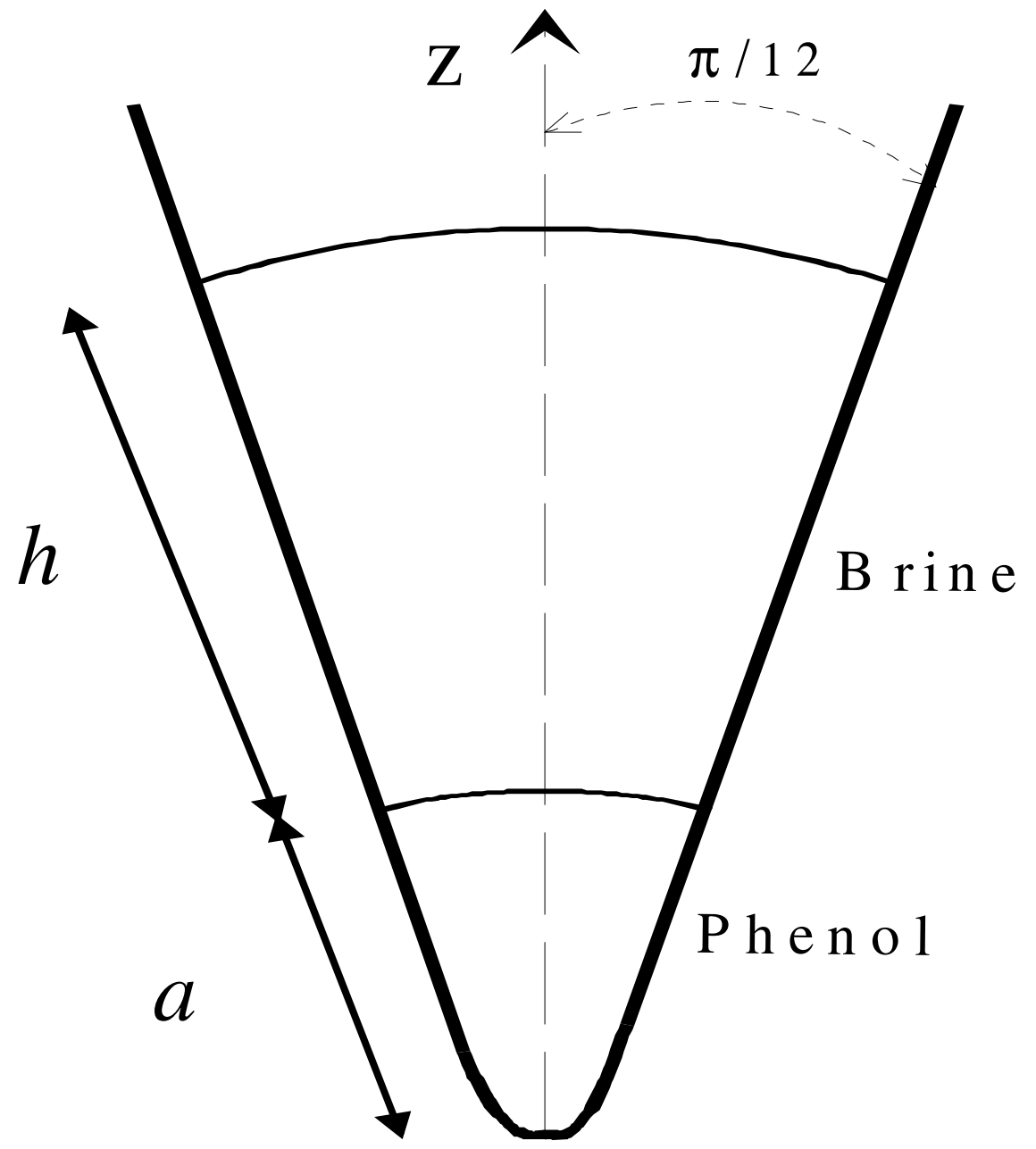

Figure 7

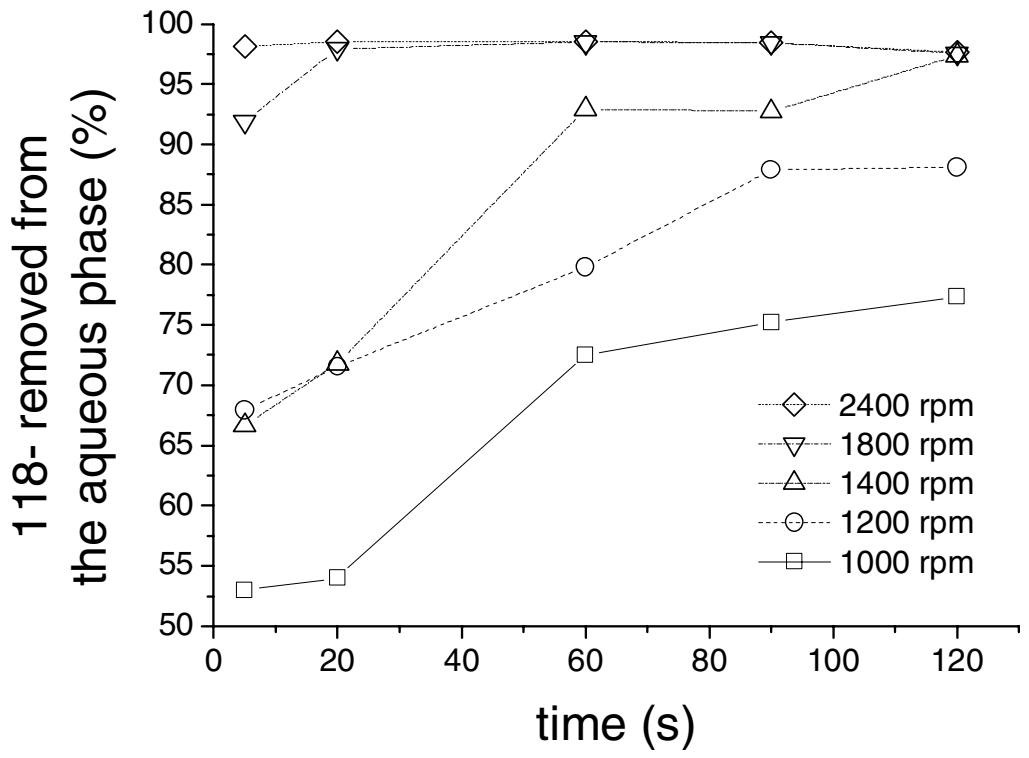

Figure 8a

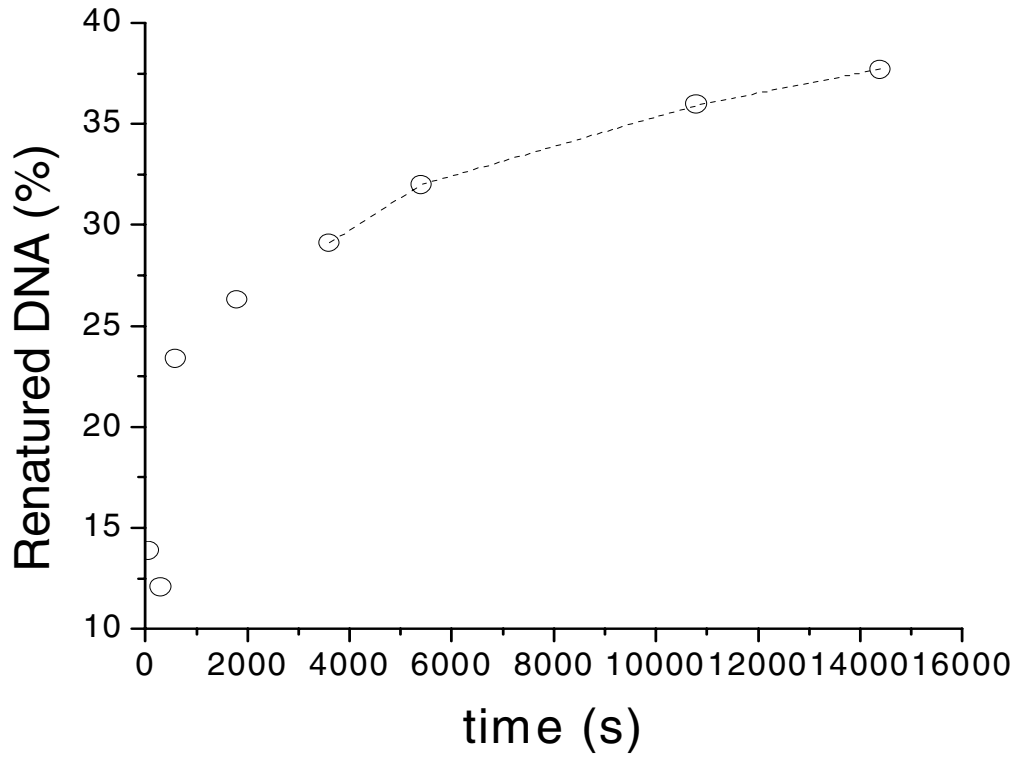

Figure 8b

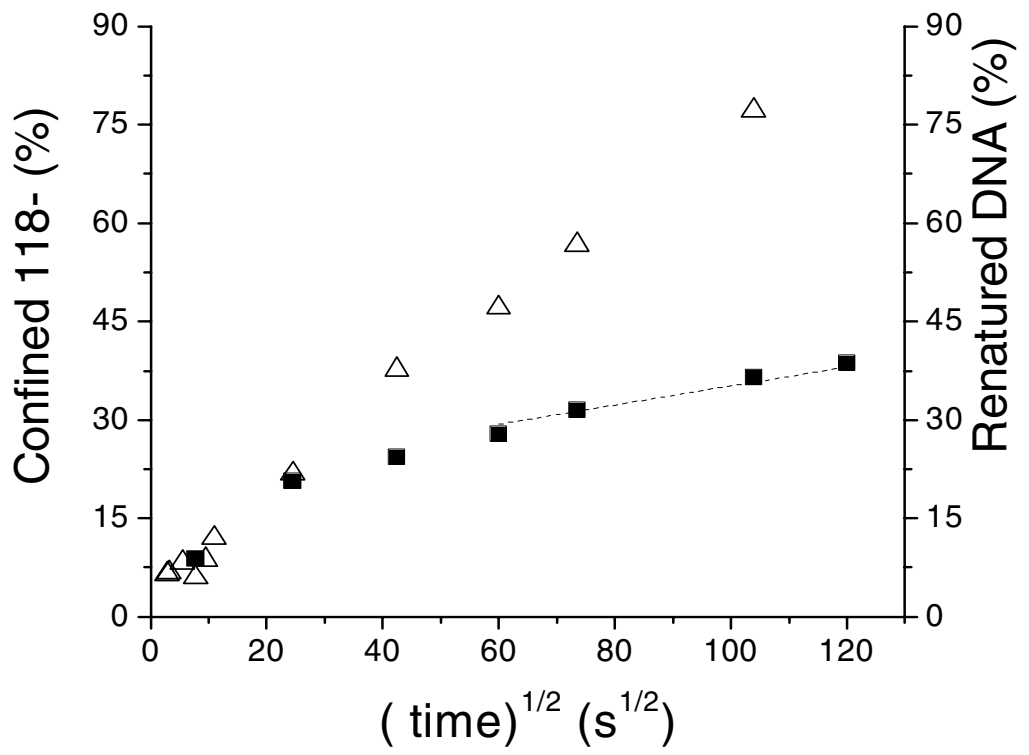

Figure 9

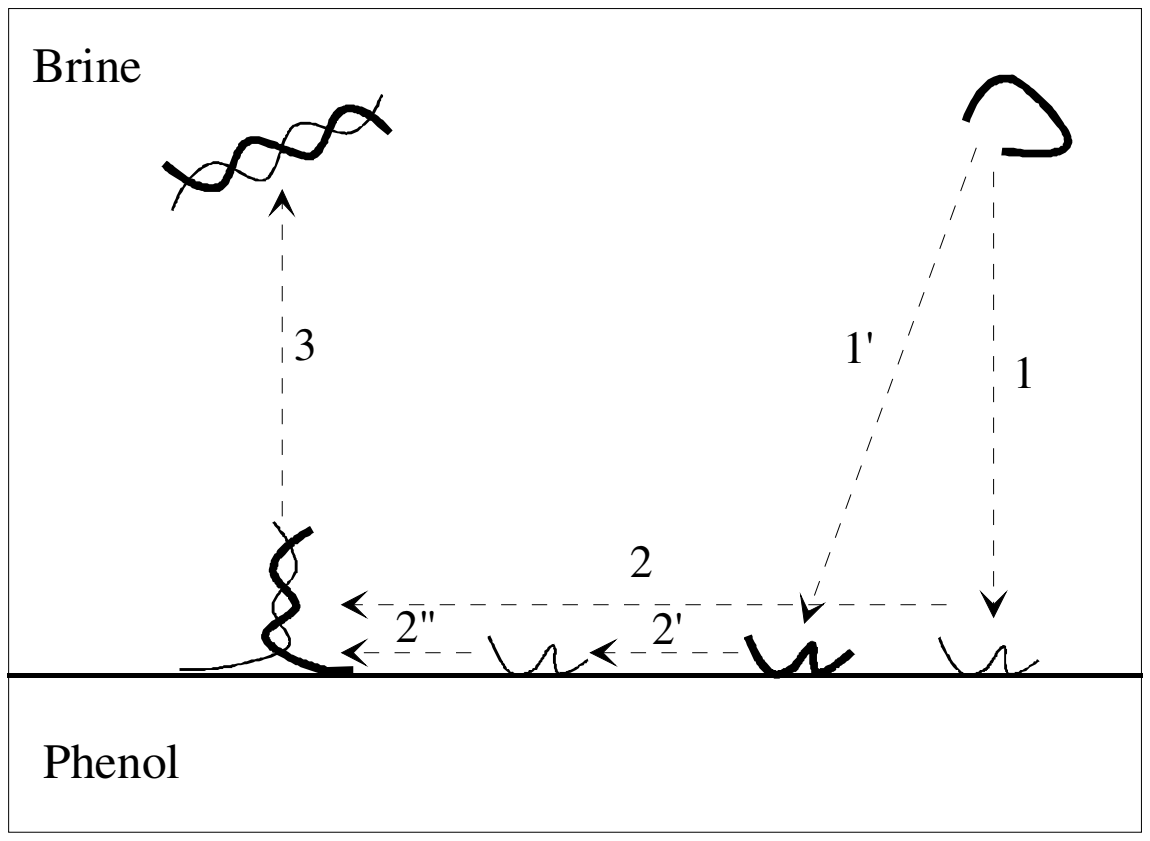

Figure 10a

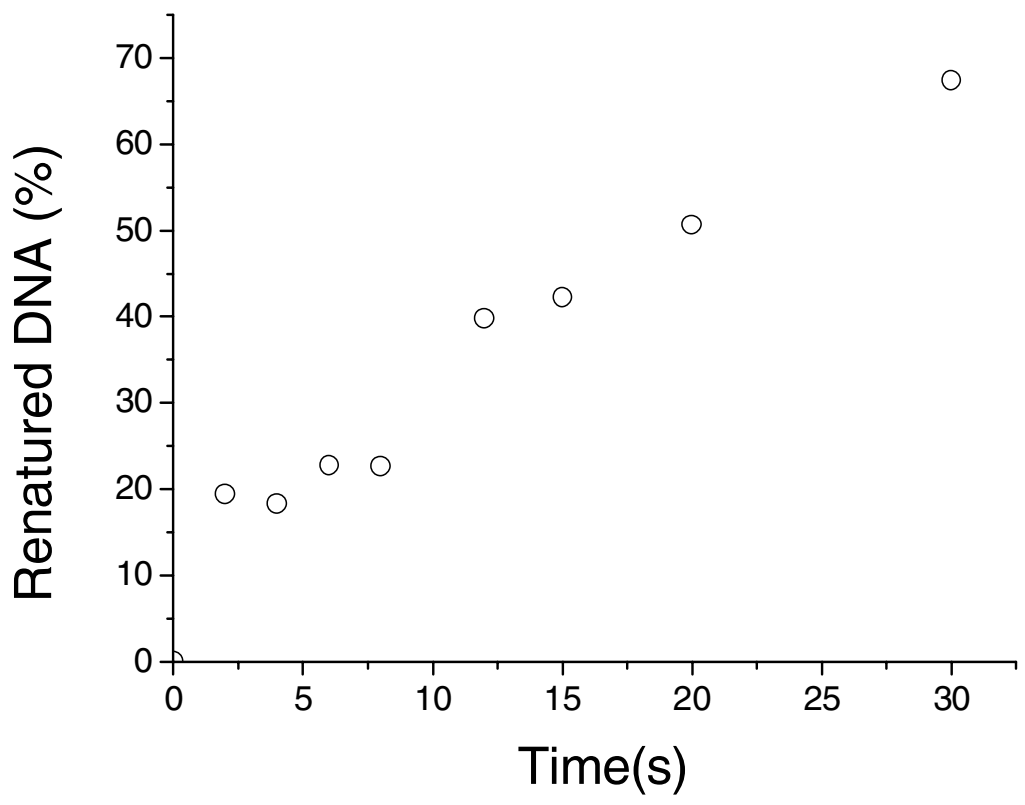

Figure 10b

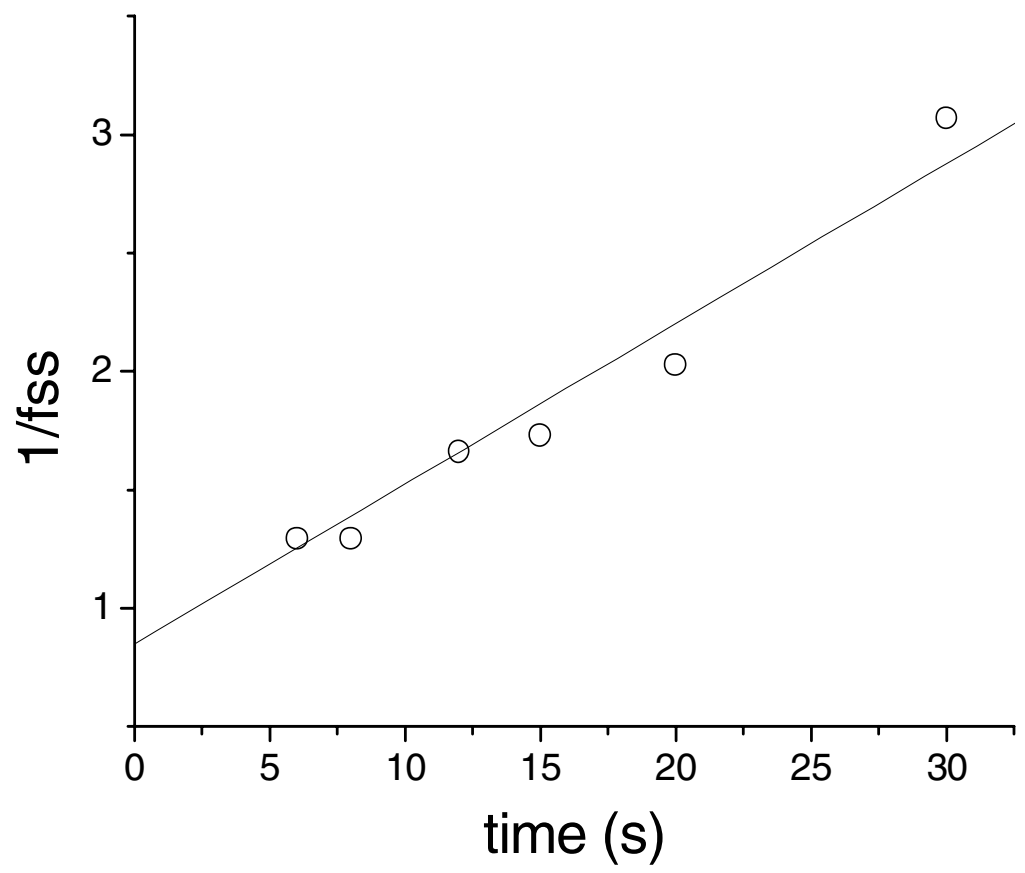

Figure 11

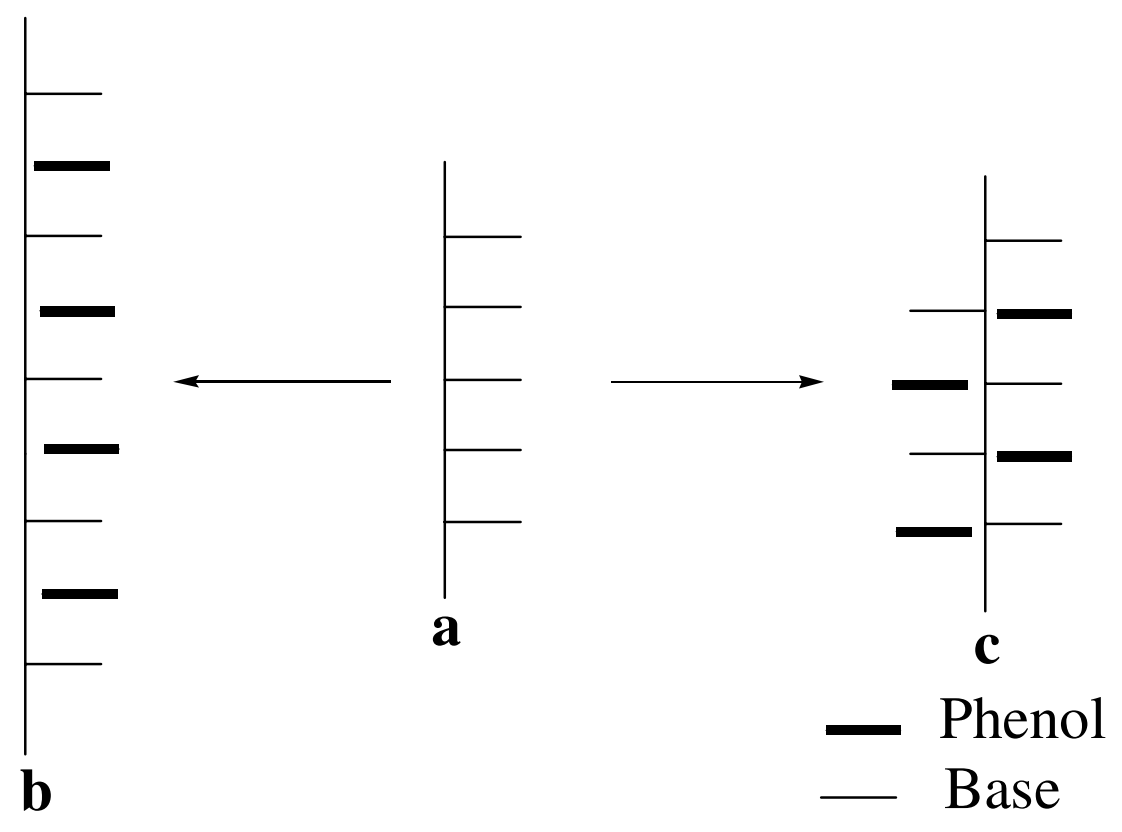

Figure 12

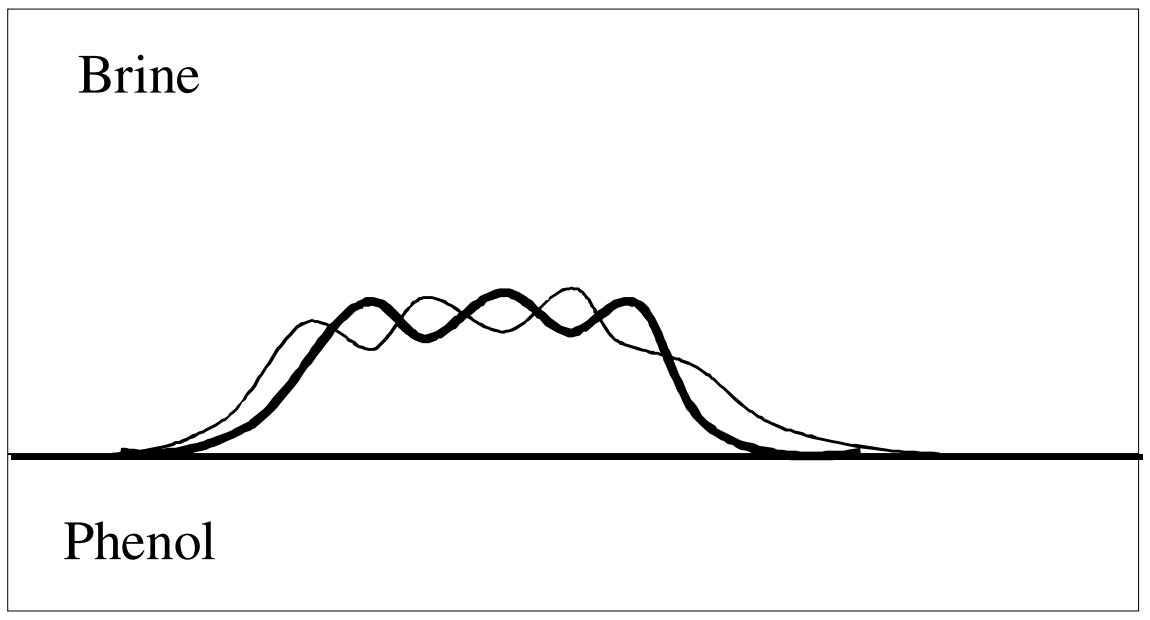